\documentclass[a4paper]{JHEP3}

\usepackage[english]{babel}
\usepackage{amsmath,amssymb}
\usepackage[dvips]{graphicx}
\usepackage{ifthen,array}
\usepackage[comma,square,sort&compress]{natbib}
\usepackage{amsfonts}
\usepackage{bbm}

\allowdisplaybreaks

\newcommand{\flux}[2][]{\ensuremath{\ifthenelse{\equal{#1}{}}{}{^{#1}\!}
\mathit{#2}}}

\newcommand{\AddrAHEP}{
  AHEP Group, Instituto de F\'{\i}sica Corpuscular --
  C.S.I.C./Universitat de Val{\`e}ncia \\
  Edificio Institutos de Paterna, Apt 22085, E--46071 Val{\`e}ncia, Spain}

 \title{Geotomography with  solar and supernova neutrinos}

\author{ E. Kh. Akhmedov\thanks{On leave from the National Research
    Centre Kurchatov Institute, Moscow, Russia} \\

The Abdus Salam International Centre for Theoretical Physics \\ 
Strada Costiera 11, 34014 Trieste, Italy \\
  E-mail:  \email{akhmedov@ictp.trieste.it}}

\author{
M.~A.~T{\'o}rtola and
  J.~W.~F.~Valle \\
  \AddrAHEP \\
  E-mail:  \email{mariam@ific.uv.es},
  \email{valle@ific.uv.es}}

\abstract{ We show how by studying the Earth matter effect on
  oscillations of solar and supernova neutrinos inside the Earth one
  can in principle reconstruct the electron number density profile of
  the Earth.  A direct inversion of the oscillation problem is
  possible due to the existence of a very simple analytic formula for
  the Earth matter effect on oscillations of solar and supernova
  neutrinos. From the point of view of the Earth tomography, these
  oscillations have a number of advantages over the oscillations of
  the accelerator or atmospheric neutrinos, which stem from the fact
  that solar and supernova neutrinos are coming to the Earth as mass
  eigenstates rather than flavour eigenstates. In particular, this
  allows reconstruction of density profiles even over relatively short
  neutrino path lengths in the Earth, and also of asymmetric profiles.
  We study the requirements that future experiments must meet to
  achieve a given accuracy of the tomography of the Earth.  }


\preprint{
IFIC/05-15 \\
  hep-ph/0502154}
  
\begin{document}

\section{Introduction}
The idea to use neutrinos in order to probe the interior of the Earth was 
originally put forward more than 30 years ago \cite{Volkova}, and since then 
has undergone a number of improvements and modifications  \cite{DeRujula:1983ya,
Wilson:1983an,Askarian:ca,Borisov:sm,Jain:1999kp,Ermilova:pw,Chechin,
Nicolaidis:fe,Nicolaidis:1990jm,Ohlsson:2001ck,Ohlsson:2001fy,
Ioannisian:2002yj,Lindner:2002wm,Winter:2005we}. In the present paper we offer 
a new insight into this problem, based on some recent theoretical and 
experimental developments in neutrino physics. 

There are in general two possible approaches to the problem of probing 
the Earth's structure with neutrinos: neutrino absorption tomography 
\cite{Volkova,DeRujula:1983ya,Wilson:1983an,Askarian:ca,Borisov:sm,Jain:1999kp} 
and neutrino oscillation tomography \cite{Ermilova:pw,Chechin,Nicolaidis:fe,
Nicolaidis:1990jm,Ohlsson:2001ck,Ohlsson:2001fy,Ioannisian:2002yj,
Lindner:2002wm}. The absorption tomography is based on the attenuation of the 
flux of neutrinos due to their scattering and absorption inside the Earth and 
so is only sensitive to the cumulative effect of neutrino absorption 
and deflection along its trajectory. Therefore the absorption tomography 
cannot give information about the matter density distribution along a given 
neutrino trajectory inside the Earth and requires many baselines to 
reconstruct the Earth's density profile.

The oscillation tomography of the Earth is possible due to the fact that 
neutrino oscillations in matter are different from those in vacuum 
\cite{Wolfenstein:1977ue,Mikheev:gs}. By studying matter effect on neutrino 
oscillations one can therefore probe the matter density distribution 
along the neutrino path. Being based on an interference phenomenon, the 
neutrino oscillation tomography has a much richer potential for studying the 
structure of the Earth. In particular, it is in principle possible to use 
just one baseline and probe the Earth's density at various points along 
the neutrino trajectory. 

In most studies on neutrino oscillation tomography, accelerators were 
considered  as the neutrino source. However, solar and supernova neutrinos 
have a number of advantages over the accelerator neutrinos in this respect 
as the probe of the Earth's interior \cite{Ioannisian:2002yj,Lindner:2002wm}. 
First and foremost, the Sun and a supernova provide us with a free source 
of neutrinos. 
In addition, it is very important that, due to the loss of coherence on 
their way to the Earth, solar and supernova neutrinos arrive at the Earth as 
mass eigenstates rather than flavour eigenstates. For oscillations of mass
eigenstate neutrinos in a medium, matter effects fully develop at much
shorter distances than they do for flavour eigenstates
\cite{Akhmedov:2000cs}, therefore by using solar or supernova
neutrinos one can probe the Earth's density distribution even over
relatively short neutrino path lengths inside the Earth.  Another
virtue of studying the oscillations of mass eigenstate neutrinos in
matter is that in that case asymmetric density profiles can be
reconstructed
\cite{Ioannisian:2002yj,deHolanda:2004fd,Ioannisian:2004jk}. 

Although to first approximation the Earth's density profile can be
considered as spherically symmetric, some deviations from perfect
symmetry are possible, especially over relatively short scales.
Exploring such short-scale inhomogeneities would be of particular
interest from the point of view of the possibility of oil or gas
prospecting \cite{Ohlsson:2001fy,Ioannisian:2002yj}.  As was shown in
\cite{Akhmedov:2001kd}, studying asymmetric density profiles through
the usual oscillations of flavour eigenstate neutrinos is practically
impossible (see also the discussion in section \ref{sec:symasym}).

Direct inversion of the neutrino oscillations problem in order to reconstruct 
the matter density profile is in general a difficult and subtle mathematical 
problem -- it reduces to the reconstruction of the potential of the 
Schr\"odinger equation from its solution. In the two-flavour case, this in 
general requires the knowledge of the energy dependence of the absolute 
value of one of the components of the neutrino wave function and also of 
the relative phase between the two components \cite{Ermilova:pw,Chechin}, 
which is not measured in neutrino flavour oscillation experiments. Therefore, 
one usually has to resort to indirect methods, for example, by generating 
random density distributions and comparing the corresponding predictions for 
oscillation probabilities with simulated data for the ``true'' profile 
\cite{Nicolaidis:1990jm,Ohlsson:2001ck,Ohlsson:2001fy}. This is a complicated 
and time consuming procedure of limited accuracy. Indeed, one normally 
represents the matter density profile along the neutrino path as a relatively 
small number of layers of constant and randomly chosen densities, which gives 
rather poor resolution due to the obvious limitation on the number of layers.

In the present paper we develop a novel {\em direct} approach to the
neutrino oscillation tomography of the Earth. It is based on a
recently found simple expression for the Earth matter effect on the
oscillations of solar and supernova neutrinos in the Earth.  The
similarity of this expression to the Fourier transform of the Earth's
density profile allows one to employ a modified inverse Fourier
transformation and reconstruct the density profile in a simple and
straightforward way.

The paper is organized as follows. In section~\ref{sec:gener} we
discuss some general features of the Earth matter effect on
oscillations of solar and supernova neutrinos inside the Earth and
present the main formulas which will be used for the inversion of the
oscillation problem. In section~\ref{sec:symasym} we consider the
advantages of oscillations of solar and supernova neutrinos over the
other neutrino oscillations in reconstructing asymmetric matter
density profiles. In section~\ref{sec:lin} we consider the inversion
procedure based on the simplest formula for the Earth regeneration
factor, valid for neutrino path length inside the Earth $L\ll 1700$ km
(linear regime). We also develop simple iteration procedures which
allow one to overcome the difficulty related to the lack of knowledge
of the regeneration factor in the domain of high neutrino energies. In
section~\ref{sec:nonlin} we briefly discuss the Earth density
reconstruction based on the more accurate expression for the Earth
regeneration factor, which allows one to study longer neutrino path
lengths in the Earth (non-linear regime).  The requirements to the
experimental setups which have to be met in order to achieve a given
accuracy of the reconstructed Earth density profile are considered in
section~\ref{sec:exptl}. In particular, we discuss the effects of
finite energy resolution of the detector on the accuracy of the
oscillation tomography of the Earth. Our main results are summarized
and discussed in section~\ref{sec:disc}. Some technical details of our
calculations are given in the Appendix.

\section{Generalities}
\label{sec:gener}

In the 3-flavour framework, neutrino oscillations are in general described
by two mass squared differences, $\Delta m_{21}^2$ and $\Delta m_{31}^2$, 
three mixing angles, $\theta_{12}$, $\theta_{13}$ and $\theta_{23}$, and
the Dirac-type CP-violating phase $\delta_{\rm CP}$. For oscillations
of solar or supernova neutrinos inside the Earth, the third mass
eigenstate essentially decouples, and the relevant parameters are
$\Delta m_{21}^2$, $\theta_{12}$ and $\theta_{13}$
\cite{Blennow:2003xw,Akhmedov:2004rq}.  The first two of these are
determined from the solar neutrino data and long-baseline reactor
experiment KamLAND \cite{solardata,KamLAND}, while for the mixing
angle $\theta_{13}$ only an upper bound exists.
For example, the recent global fit of neutrino 
data of ref.~\cite{Maltoni:2004ei} gives for the solar neutrino oscillation 
parameters the $3\sigma$ allowed ranges $\theta_{12}=(28.7\div 38.1)
^\circ$ and $\Delta m_{21}^2=(7.1\div 8.9)\times 10^{-5}$ eV$^2$, 
with the best-fit values $\theta_{12}=33.2^\circ$ and $\Delta m_{21}^2 =
7.9\times 10^{-5} $eV$^2$, while for the mixing angle $\theta_{13}$ 
one finds $\theta_{13} \lesssim 9.1^\circ~(13.1)^\circ$, or $\sin \theta_{13} 
\lesssim 0.16~(0.23)$ at 90\% C.L. (3$\sigma$). 

Consider a flux of solar or supernova neutrinos arriving at the Earth 
and traveling a distance $L$ inside the Earth before reaching the 
detector. Due to the loss of coherence on their way to the Earth,
the incoming neutrinos represent an incoherent sum of fluxes of 
mass-eigenstate neutrinos (see, e.g., ref.~\cite{Dighe:1999id}).
The Earth matter effect on oscillations of such neutrinos inside the 
Earth is fully described by the so-called regeneration factor 
$P_{2e}^{\oplus}-P_{2e}^{(0)}$. Here $P_{2e}^{\oplus}$ is the probability 
that a neutrino arriving at the Earth as a mass eigenstate $\nu_2$ is 
found at the detector in the $\nu_e$ state after having traveled a 
distance $L$ inside the Earth, and $P_{2e}^{(0)}$ is the projection of the 
second mass eigenstate onto $\nu_e$: $P_{2e}^{(0)}=|U_{e2}|^2$, $U$ being 
the leptonic mixing matrix in vacuum. Note that $P_{2e}^{(0)}$ is in fact 
the value of $P_{2e}^\oplus$ in the limit of vanishing matter density  
or zero distance traveled inside the Earth.

As has been shown in \cite{Akhmedov:2004rq}, in the 3-flavour 
framework the Earth 
regeneration factor for solar and supernova neutrinos can be written as 
\begin{equation}
P_{2e}^\oplus-P_{2e}^{(0)}=\frac{1}{2}\,\cos^2\theta_{13}\,\sin^2 
2\theta_{12}\,f(\delta)\,,
\label{reg}
\end{equation}
where
\begin{equation}
f(\delta)=\int_0^L\! dx \,V(x) \sin\left[2\int\limits_x^{L}\!\omega(x')
\,dx'\right]\,, 
\label{f1}
\end{equation}
with
\begin{equation}
\omega(x)=\sqrt{[\cos 2\theta_{12}\,\delta-V(x)/2]^2+\delta^2
\sin^2 2\theta_{12}}\,,\qquad\qquad 
\delta=\frac{\Delta m_{21}^2}{4E}\,. 
\label{omega}
\end{equation}
The effective matter-induced potential of neutrinos $V(x)$ in eqs. 
(\ref{f1}) and (\ref{omega}) is related to the charged-current 
potential 
$V_{\rm CC}(x)$ through
\begin{equation}
V(x)=\cos^2\theta_{13} V_{\rm CC}(x)=\cos^2\theta_{13} \sqrt{2}\,G_F\,
N_e(x)\,,
\label{Vcc}
\end{equation}
where $G_F$ is the Fermi constant and $N_e(x)$ is the electron number density 
in matter, $x$ being the coordinate along the neutrino path in the Earth. 
The 2-flavour ($\theta_{13}=0$) version of eqs.~(\ref{reg})-(\ref{omega}) 
was derived in~\cite{Ioannisian:2004jk}, and similar formulas were also 
found in ref.~\cite{deHolanda:2004fd}.  

Equations (\ref{reg}) and (\ref{f1}) were obtained under the assumption 
$V(x)\ll 2\delta$, which is very well satisfied for oscillations of solar 
neutrinos in the Earth. It is also satisfied with a good accuracy for 
supernova neutrinos (except for very high energy ones, which are on the 
tail of the supernova neutrino spectrum).  If, in addition, one also 
requires $V L\ll 1$, eq.~(\ref{f1}) simplifies to \cite{Akhmedov:2004rq} 
\begin{equation}
f(\delta)=\int_0^L V(y)\sin 2\delta (L-y) dy\,.
\label{f2}
\end{equation}

Equation (\ref{f2}) is very suggestive: it has a Fourier integral 
form and actually means that in the small $V$ limit the function 
$f(\delta)$ is just the Fourier transform of matter-induced neutrino 
potential $V(x)$ 
\footnote{The finite range of integration in (\ref{f2}) is related to 
the fact that the function $V(x)$ is defined on the finite interval 
$0\le x \le L$.}.
Therefore, if $f(\delta)$ is determined experimentally through 
eq.~(\ref{reg}), one can employ the inverse Fourier transformation to 
reconstruct the effective matter-induced potential  $V(x)$: 
\begin{equation}
V(x)=\frac{4}{\pi}\int_0^\infty f(\delta)\sin 2\delta (L-x) d\delta\,.
\label{inv1}
\end{equation}
The electron number density profile of the Earth $N_e(x)$ can then be 
found from eq.~(\ref{Vcc}). 

The Earth density profile could in principle be exactly reconstructed 
from the solar and supernova neutrino data through eq. (\ref{inv1}) 
under the following conditions: 

\begin{enumerate}
\item
Eqs. (\ref{reg}) and (\ref{f2}) are exact;

\item
The function $f(\delta)$ is precisely measured in the whole interval 
$0\le\delta < \infty$ (i.e. in the infinite interval of neutrino energies 
$0\le E < \infty$);

\item The $\delta$-dependence of the function $f(\delta)$ is known
  precisely, i.e the detectors have perfect energy resolution, and can
  determine the energy of incoming neutrinos from those of the
  secondary particles. In addition, the neutrino parameters $\Delta
  m_{21}^2$, $\theta_{12}$ and $\theta_{13}$ are precisely known.
\end{enumerate}

In reality, none of these conditions is satisfied: eqs. (\ref{reg}) and 
(\ref{f2}) are only valid in the limit $V(x)\ll 2\delta$, $\,V L\ll 1$, 
the regeneration factor $P_{2e}^\oplus-P_{2e}^{(0)}$ (and so the function 
$f(\delta))$ can only be measured in a finite interval of energies and with 
some experimental errors, 
the detectors have finite energy resolution and can only 
give limited information on the energy of incoming neutrinos, and 
the neutrino parameters are only known with certain experimental 
uncertainties. In what follows we will study the constraints that these 
limitations put on the accuracy of the reconstructed potential $V(x)$, 
by relaxing conditions (1) -- (3) one by one. Conversely, we shall discuss 
the requirements that are put on the experimental installations by the 
condition of reconstructing the potential $V(x)$ with a given accuracy. 
We will also discuss the ways in which some of the above-mentioned 
limitations can be overcome. 

\section{Symmetric versus asymmetric density profiles}
\label{sec:symasym}

By symmetric density profiles we mean the profiles that are symmetric with 
respect to the midpoint of the neutrino trajectory inside the Earth. 
They give rise to the potentials that have the same property, i.e. 
\begin{equation}
V(L-x)=V(x)\,. 
\label{sym}
\end{equation}
If the electron number density of the Earth $N_e$ was 
exactly spherically symmetric, the corresponding neutrino potential $V(x)$ 
would have satisfied eq.~(\ref{sym}). However, this symmetry is only 
approximate; in particular, it is violated by inhomogeneities of the Earth's 
density distribution on short length scales. Studying these inhomogeneities  
may be especially interesting, e.g., from the point of view of possible oil or 
gas prospecting. 
        
Effects of asymmetric density profiles on oscillations of solar
neutrinos in the Earth have been previously discussed in
\cite{Ioannisian:2002yj,deHolanda:2004fd, Ioannisian:2004jk}. Here we
give a more detailed discussion of neutrino oscillations in asymmetric
matter and also compare in this context mass-to-flavour and pure
flavour neutrino oscillations.

It is easy to show that in the two-flavour (2f) framework the probabilities 
of oscillations between neutrinos of different flavour are the same for the
potentials $V(x)$ and \mbox{$V(L-x)$}. Indeed, the 2f neutrino oscillation 
probabilities are invariant under the time reversal transformation $P_{ab}\to 
P_{ba}$. This follows from the unitarity relations 
\begin{eqnarray}
P_{aa}+P_{ab} &=& 1\,, \nonumber \\
P_{aa}+P_{ba} &=& 1\,,
\label{unit1}
\end{eqnarray}
which enforce $P_{ab}=P_{ba}$ \cite{deGouvea:2000un,Akhmedov:2001kd}. On the 
other hand, for an arbitrary number of flavours, time reversal transformation 
of the probabilities of neutrino oscillations in matter is equivalent to 
flipping the sign of the Dirac-type CP-violating phases $\{\delta_{\rm CP}\}$ 
and replacing the potential $V(x)$ with the reverse potential $V(L-x)$ 
\cite{Akhmedov:2001kd}. Since Dirac-type CP-violation is absent in the 2f 
case, from $P_{ab}=P_{ba}$ one immediately finds that 2f oscillation 
probabilities are invariant under the transformation $V(x)\to V(L-x)$. 

This can also be expressed in the following way. The evolution matrix $S$ of 
a 2f neutrino system is a $2\times 2$ unitary matrix which 
can be written in the flavour eigenstate basis as 
\begin{equation} 
S=\left( \begin{array}{cc} ~\alpha & ~\beta
\\ -\beta^* & ~~\alpha^* \end{array} \right) 
\label{S} 
\end{equation}
with $|\alpha|^2+|\beta|^2=1$. In terms of the elements of $S$, the 
oscillation probabilities are given as $P_{ab}=|S_{ba}|^2$. Hence, time 
reversal of the evolution matrix (which in the 2f case is equivalent to the 
transformation $V(x)\to V(L-x)$) reduces to the transposition $S_{ab}\to 
S_{ba}$, i.e.  
\begin{equation} 
\alpha \to \alpha\,, \qquad\qquad \beta \to -\beta^*\,. 
\label{Trev} 
\end{equation}
Since the transition probabilities $P_{ab}=|\beta|^2$, as well as the survival 
probabilities $P_{aa}=|\alpha|^2$, are invariant under the transformation 
(\ref{Trev}), they cannot discriminate between the potentials $V(x)$ and 
$V(L-x)$. This means that flavour oscillations cannot be used for a unique   
reconstruction of asymmetric density profiles: there will always be a two-fold 
ambiguity.

The situation is drastically different when one considers the transitions 
between the mass and flavour eigenstates, as is the case for oscillations 
of solar and supernova neutrinos inside the Earth. In that case the 
unitarity conditions read
\begin{eqnarray}
P_{1e} \,&+& \, P_{2e} = 1\,, \nonumber \\
P_{1e}^{(0)} &+& P_{2e}^{(0)} = 1\,,
\label{unit2}
\end{eqnarray}
from which one only finds $P_{2e}-P_{2e}^{(0)}=-(P_{1e}^{(0)}-P_{1e}^{(0)})$, 
and no restrictions on the behaviour of the probabilities under the 
replacement $V(x)\to V(L-x)$ follow. In fact, it was shown in 
\cite{Akhmedov:2004rq} that in the 2f case the Earth regeneration factor 
$P_{2e}-P_{2e}^{(0)}$ can be expressed through the elements of the matrix 
$S$ as
\begin{equation}
P_{2e}-P_{2e}^{(0)}=\cos 2\theta_{12} |\beta|^2+2 \sin 2\theta_{12}{\rm Re}
(\alpha^* \beta)\,.
\label{TT}
\end{equation} While the first term on the right-hand side of eq.~(\ref{TT}) is
invariant with respect to the transformation (\ref{Trev}), the second
is in general not \footnote{It is only invariant when $\beta$ is pure
  imaginary, which is the case for symmetric density profiles
  \cite{Akhmedov:2001kd}. }.  Thus, unlike the flavour oscillations,
the oscillations of mass eigenstate neutrinos into flavour eigenstate
ones can be used to uniquely reconstruct asymmetric density profiles
even in the 2f framework. As was shown in
refs.~\cite{Blennow:2003xw,Akhmedov:2004rq}, in the case of
oscillations of solar or supernova neutrinos inside the Earth the
third neutrino flavour essentially decouples, and to a very high
accuracy the problem is reduced to an effective 2-flavour one.
Therefore, our conclusion about the impossibility of using the neutrino
flavour oscillations inside the Earth for an unambiguous
reconstruction of asymmetric density profiles holds also in the
  3-flavour case, hence the superiority of mass-to-flavour
  oscillations.  This is in accord with the previous findings that
matter-induced T violation in neutrino flavour oscillations inside the
Earth is too small to be measured~\cite{Akhmedov:2001kd}.

\section{Linear regime}
\label{sec:lin}

The linear regime of the inverse problem of neutrino oscillations in the Earth is 
based on the simple formula (\ref{f2}) for the function $f(\delta)$, which was 
derived under the assumptions
\begin{equation}
V/2\delta \ll 1\,,\qquad\quad V L\ll 1\,. 
\label{cond2}
\end{equation} 
For the matter density in the upper mantle of the Earth,
$\rho\simeq 3$ g/cm$^3$, the second of these conditions leads to the
upper limit on the allowed neutrino path lengths in the Earth \begin{equation} L\ll
1700 ~\mbox{km}\,,
\label{upper}
\end{equation}
which we will now assume to be satisfied. This condition will be relaxed in 
the discussion of the non-linear regime in section~\ref{sec:nonlin}. 
\subsection{Integration over finite energy intervals}
\label{sec:finiteInt}

We shall first assume that eqs. (\ref{reg}) and (\ref{f2}) are exact, 
but the function $f(\delta)$ is only known in a finite interval 
$[\delta_{min}, \delta_{max}]$ (i.e. in a finite interval of neutrino 
energies $E_{min}\le E\le E_{max}$). To study the effects of finite 
$\delta_{min}$ and $\delta_{max}$, consider the integral of the type 
(\ref{inv1}) in the finite limits:
\begin{equation}
\frac{4}{\pi}\int_{\delta_{min}}^{\delta_{max}}\!f(\delta)\sin 2\delta 
(L-x) d\delta=\frac{1}{\pi}\int_0^L\!dy V(y)\left.\!\left\{\frac{\sin 
2\delta (x-y)}{x-y}-\frac{\sin 2\delta(2L-x-y)}{2L-x-y}\right\}
\right|_{\delta_{min}}^{\delta_{max}}.
\label{aux1}
\end{equation}
Here we have used eq. (\ref{f2}), changed the order of integrations 
and performed the integral over $\delta$.

Ideally, one would like to have $\delta_{min} L\ll 1$ and $\delta_{max} L\gg 1$ 
in order that the integral in eq.~(\ref{aux1}) approach the integral over the 
infinite interval $0\le \delta < \infty$ in eq.~(\ref{inv1}) as closely as 
possible. As we shall see, having large enough $\delta_{max}$ in principle does 
not pose a problem.
In contrast to this, in most situations of practical interest $\delta_{min}
\gtrsim L^{-1}$, i.e. the condition $\delta_{min} L\ll 1$ is not satisfied, 
which could be a serious problem. 
We shall show, however, that this difficulty can be readily overcome.  

\begin{figure}[htbp]
  \centering
\includegraphics[height=7.5cm,width=.7\linewidth,clip]{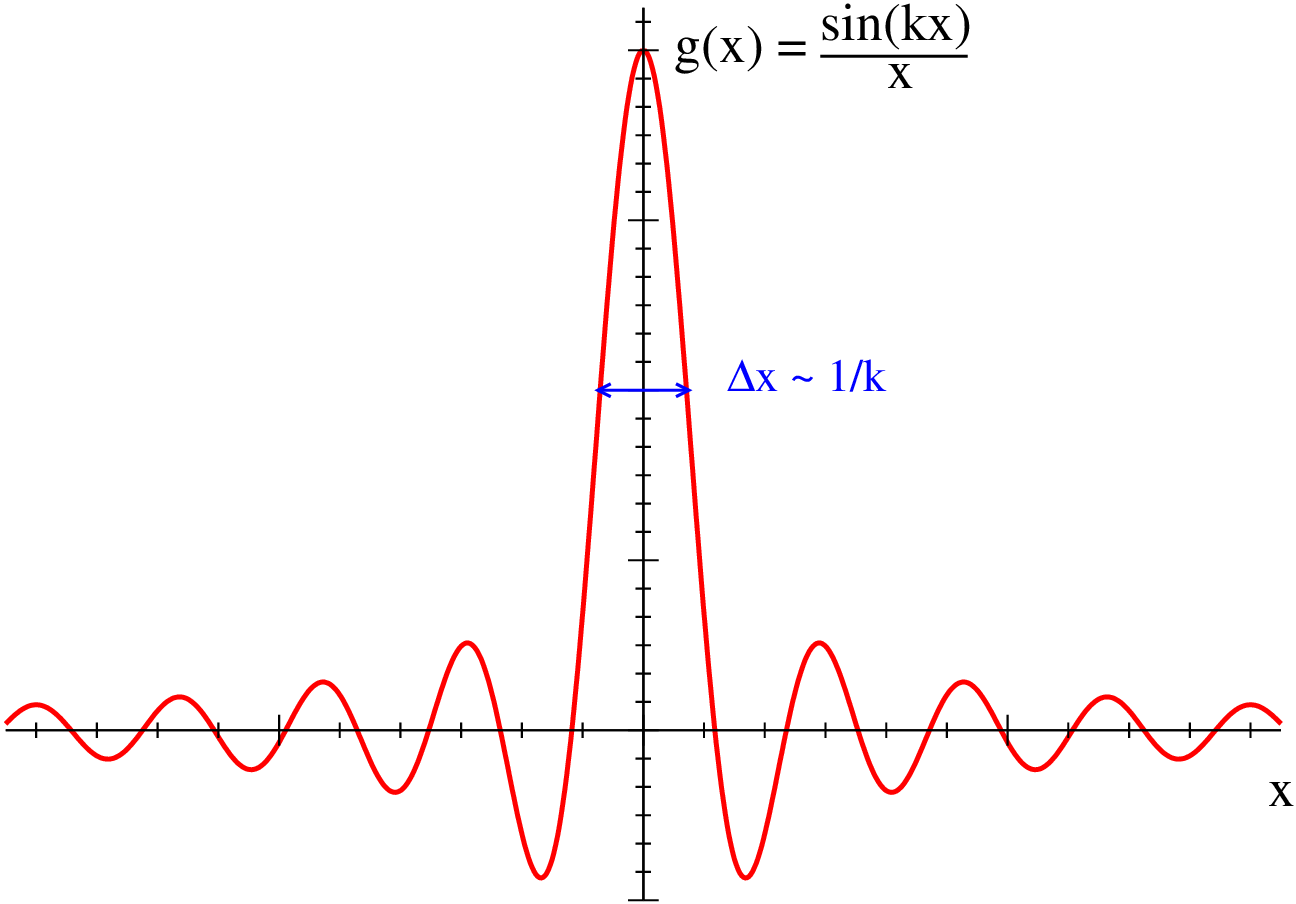}
  \caption{Function $g(x)=\sin(kx)/x$}
 \label{fig:g}
\end{figure}

Let us first study the effect of finite $\delta_{max}$ in eq.~(\ref{aux1}). 
In the limit $k\to\infty$, the function $\sin kx/x$ goes to $\pi\delta(x)$ 
\footnote{Note that here $\delta(x)$ denotes Dirac's delta-function, not to 
be confused with the parameter $\delta$ defined in eq. (\ref{omega}).}, 
therefore for $\delta_{max}\to \infty$ the upper limit of the integral in 
(\ref{aux1}) would yield $V(x)-V(2L-x)$, i.e. essentially the potential $V(x)$ 
(note that $V(2L-x)=0$ for all $x\ne L$). The function $g(x)=\sin kx/x$ for 
finite $k>0$ is plotted in fig.~\ref{fig:g}. The width of its central peak is 
$\simeq \pi/k$. With increasing $k$, the peak becomes higher and narrower, and 
the amplitude of the side oscillations quickly decreases. It is therefore 
clear that finite $\delta_{max}$ leads to the finite coordinate resolution 
$\Delta x\simeq \pi/\delta_{max}$ of the reconstructed potential $V(x)$ as 
well as to small oscillations of the reconstructed potential around the 
true one. 

Large values of $\delta$ correspond to small values of neutrino energy 
$E$ (see eq.~(\ref{omega})), therefore, the smaller the neutrino energy, 
the shorter the coordinate scale on which the Earth density profile can 
be probed. This may look somewhat counter-intuitive; however, one 
should remember that we are probing the Earth density distribution with 
neutrino oscillations rather than with direct neutrino-matter 
interactions. The smaller the neutrino energy, the shorter the 
oscillation length, and so the finer the density structures that can 
be probed. 
In future low-energy solar neutrino experiments, neutrinos of energies
as small as a hundred keV  to 1 MeV can probably be detected
\cite{Raghavan,Ejiri:1999rk,LOWE}; this would correspond to
$\delta_{max}\simeq (2\div 20)\times 10^{-11}$ eV, or $\Delta x\sim
\pi/\delta_{max}\simeq (3\div 30)$ km, which is a very good coordinate
resolution \footnote{This holds in the idealized case of perfect
  energy resolution of neutrino detectors. Finite energy resolution of
  the detectors is expected to reduce the coordinate resolution of the
  reconstruction procedure, see section~\ref{sec:finiteE}.}. For the
neutrino path lengths $L\gtrsim 100$ km one finds $\delta_{max} L
\gtrsim (10\div 100)$, i.e. the condition $\delta_{max} L \gg 1$ can
be easily met.

At the same time, as we have already mentioned, having a sufficiently
small $\delta_{min}$ may be a fundamental problem. There are two
reasons for that.  First, small $\delta_{min}$ implies large neutrino
energies, and there are upper limits to the available neutrino
energies. The spectrum of solar neutrinos extends up to 15 MeV ($hep$
neutrinos have slightly higher energies, but their flux is very
small); the average energy of supernova neutrinos is $\sim 20$
MeV~\cite{Keil:2002in}.  Yet, this problem can to some extent be
alleviated, at least in principle.  It might be possible to measure
the Earth matter effect for supernova neutrinos of energies up to
$\sim 100$ MeV, which are on the high-energy tail of the spectrum, but
still not too far from the mean energy. Thus, for a nearby supernova
and large enough detectors one can probably have sufficient
statistics.

The second obstacle is of more fundamental nature. Our expressions (\ref{f1}) 
and (\ref{f2}) are only valid in the approximation $V/2\delta \ll 1$, which 
may break down for too small $\delta_{min}$. This gives a lower limit on the 
values of $\delta_{min}$ one can use, depending on the accuracy of the 
reconstructed potential one wants to achieve. In principle, one could attempt 
at deriving a more general expression for $f(\delta)$, not relying on the 
approximation $V/2\delta \ll 1$; in particular, it is fairly easy to study the 
opposite case $V/2\delta \gg 1$. However, for $V\gtrsim 2\delta$ the 
expression for $f(\delta)$ does not have a simple dependence on the 
potential $V(x)$, and solving the inverse problem of neutrino oscillations 
becomes a difficult task. 

For an estimate of the constraint on $\delta_{min}$ imposed by the condition 
of small $V/2\delta$, let us require $V/2\delta_{min} < 1/5$. For the matter 
density in the upper mantle of the Earth, $\rho\simeq 3$ g/cm$^3$, this gives 
$\delta_{min}\gtrsim 2.8\times 10^{-13}$ eV, which for $\Delta m_{21}^2 
\simeq 7.9\times 10^{-5}$ eV$^2$ leads to $E_{max}\lesssim 70$ MeV. For 
neutrinos passing through the core of the Earth, the constraints on 
$\delta_{min}$ and $E_{max}$ will be a factor of 3 -- 4 more stringent.  

Let us now estimate the magnitude of $\delta_{min}L$ corresponding to $L\simeq 
300$ km, which is a representative value of neutrino potholing inside the Earth, 
satisfying condition (\ref{upper}). For solar neutrinos ($E_{max}
\simeq 15$ MeV) we find $\delta_{min}L\simeq 2$, while for supernova neutrinos, 
taking $E_{max}\simeq 70$ MeV, we obtain $\delta_{min}L\simeq 0.43$. 
Thus, except for very small values of $L$, we have to deal with situations 
when $\delta_{min}L\gtrsim 1$. 
 
How large is the error introduced by non-vanishing $\delta_{min}$ in
the reconstruction of the density profile $V(x)$?  To study that, let
us consider the integral of the type (\ref{inv1}) with the finite
lower integration limit $\delta_{min}= L^{-1}$.  In
fig.~\ref{fig:step-dmin1} the lowest curve gives the result of such a
calculation for the step-function density profile (shown by the dashed
line) and $\delta_{max}=300 L^{-1}$. One can see several interesting
features of the result. First, the deviation from the exact profile is
relatively small near the detector ($x\simeq L$), but reaches about a
factor of three far from it ($x\simeq 0$). Second, despite a
significant deviation from the exact profile $V(x)$, the positions and
the magnitudes of the jumps in $V(x)$ are reproduced very accurately.
Both these features can be easily understood (see sections
\ref{sec:iterI} and \ref{sec:small} below).

Thus, we have seen that the error in the reconstructed profile due to the lack 
of the knowledge of the Earth matter effect in the domain of low $\delta$ (high 
energies) can be quite substantial. We shall show now how this problem can be 
cured by invoking simple iteration procedures.

\subsection{Iteration procedure I}
\label{sec:iterI}

Let us study the effect of non-vanishing $\delta_{min}$ in more detail. In 
order to do so, we consider the limit $\delta_{max}\to \infty$, which is 
justified by the preceding discussion. From eq.~(\ref{aux1}) one then readily 
finds 
\begin{equation}
V(x)=\frac{4}{\pi}\int_{\delta_{min}}^\infty f(\delta)\sin 2\delta (L-x) 
d\delta +\frac{1}{\pi}\int_0^L V(y) F(x,y; 2\delta_{min}) dy\,,
\label{inv2}
\end{equation}
where the function $F(x,y;k)$ is defined as  
\begin{equation}
F(x,y;k)=\frac{\sin k(x-y)}{x-y}-\frac{\sin k(2L-x-y)}{2L-x-y}\,.
\label{F1}
\end{equation}
It is symmetric with respect to its first two arguments: $F(x,y; k)=F(y,x; k)$. 

Equation (\ref{inv2}) is exact provided that eq. (\ref{f2}) is exact. 
By comparing it with eq.~(\ref{inv1}), we find that the second integral in 
(\ref{inv2}) can be considered as compensating for an error introduced in 
eq. (\ref{inv1}) by having a non-zero lower limit in the integral over 
$\delta$. However, this compensating integral cannot be calculated directly 
because it contains the unknown potential $V(x)$. Thus, we have traded 
one unknown quantity -- the function $f(\delta)$ in the domain $\delta < 
\delta_{min}$ -- for another.
At first sight, this does not do us any good. This is, however, incorrect:
eq.~(\ref{inv2}) allows a simple iterative solution.

We first note that in the limit $\delta_{min}\to 0$ the second integral in 
eq.~(\ref{inv2}) disappears, while the first one yields $V(x)$. Therefore, for 
not too large values of $\delta_{min}$ the first term in (\ref{inv2}) is 
expected to give a reasonable first approximation to $V(x)$. One can then use 
the result in the second integral to obtain the next approximation to $V(x)$, 
and so on. Thus, we define
\begin{eqnarray}
V_0(x) &=& \frac{4}{\pi}\int_{\delta_{min}}^\infty f(\delta)\sin 
2\delta (L-x) d\delta\,, 
\label{V0m}\\ 
I_0(x) &=&\frac{1}{\pi}\int_0^L V_0(y) F(x,y; 2\delta_{min}) dy\,,
\qquad\quad V_1(x)=V_0(x)+I_0(x)\,, 
\label{I0}
\\
& & \nonumber \\
& & \dots\dots \nonumber \\
& & \nonumber \\
I_{n-1}(x) &=&\frac{1}{\pi}\int_0^L V_{n-1}(y) F(x,y; 2\delta_{min}) dy\,,
\qquad\quad \!\!V_n(x)=V_0(x)+I_{n-1}(x)\,.
\label{iter1}
\end{eqnarray}
This gives a sequence of potentials $V_0(x),\,V_1(x),\dots,\,V_n(x),
\dots$ which, for small enough $\delta_{min}$, converges to $V(x)$. It 
should be noted that, while the exact eq.~(\ref{inv2}) is independent 
of $\delta_{min}$,\footnote{Its left-hand side is $\delta_{min}$-independent, 
so must be the right-hand side. Actually, by {\em requiring} that the 
derivative of the right-hand side of (\ref{inv2}) with respect to 
$\delta_{min}$vanish, one can recover eq.~(\ref{f2}) for $f(\delta)$.} 
our iteration procedure involves 
the approximate potentials and so depends on it. The smaller 
the chosen value of $\delta_{min}$, the faster the convergence of $V_n(x)$ 
to $V(x)$; for $\delta_{min}$ exceeding some critical value (which in 
general depends on the profile $N_e(x)$) the iteration procedure fails.

\begin{figure}[htbp]
  \centering
  \includegraphics[width=.9\linewidth,clip]{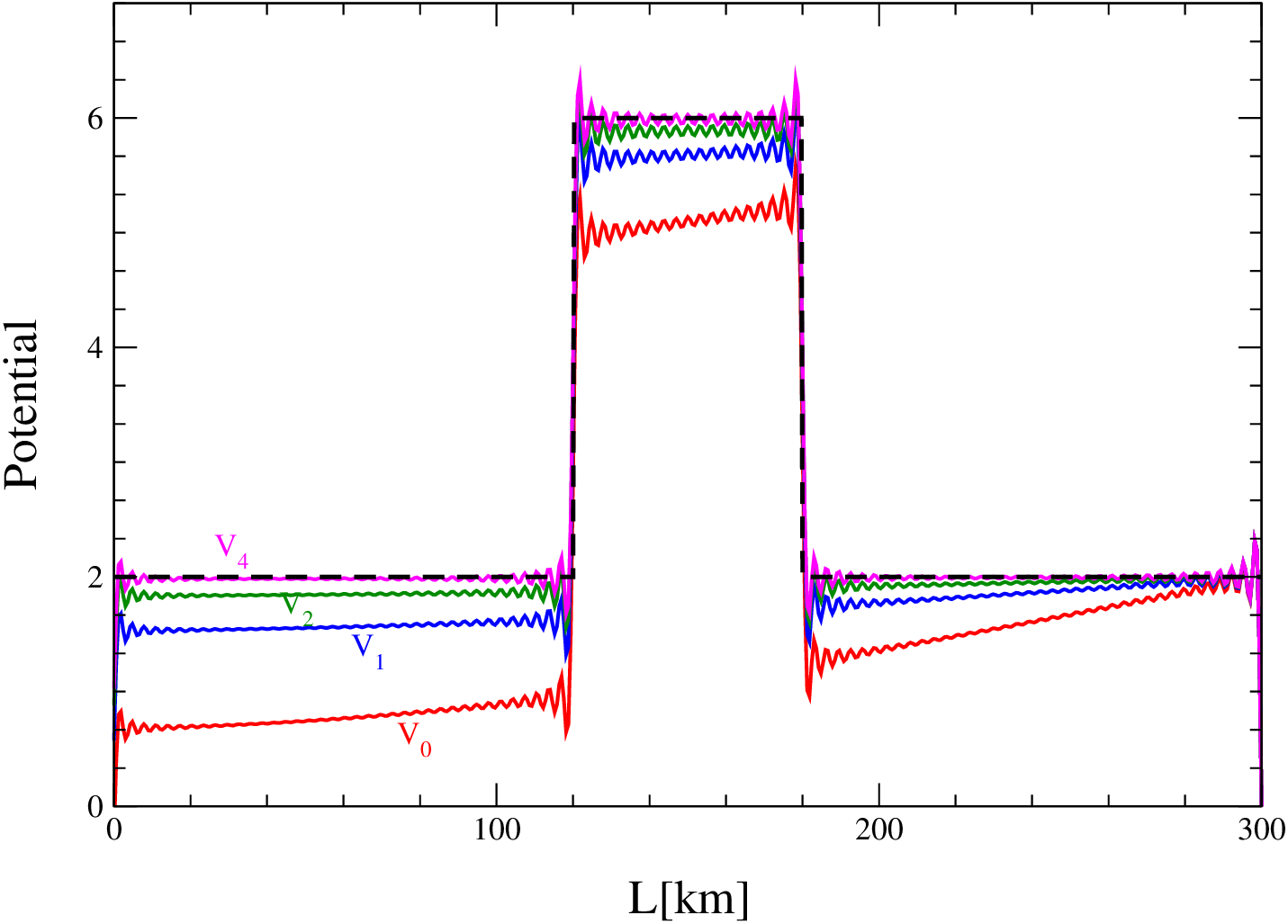}
  \caption{Step-function potential $V(x)$ (dashed line), zeroth order 
   reconstructed potential $V_0(x)$ and the results of the first, second and 
   fourth iterations $V_1(x)$, $V_2(x)$ and $V_4(x)$. The vertical scale is 
   that of the corresponding matter density in g/cm$^3$, assuming $Y_e=0.5$ and 
   $\theta_{13}=0$. The following values of the parameters were chosen: 
   $\delta_{min}=L^{-1}$, $\delta_{max}=300 L^{-1}$.
   } 
 \label{fig:step-dmin1}
\end{figure}

\begin{figure}[htbp]
  \centering
  \includegraphics[width=.9\linewidth,clip]{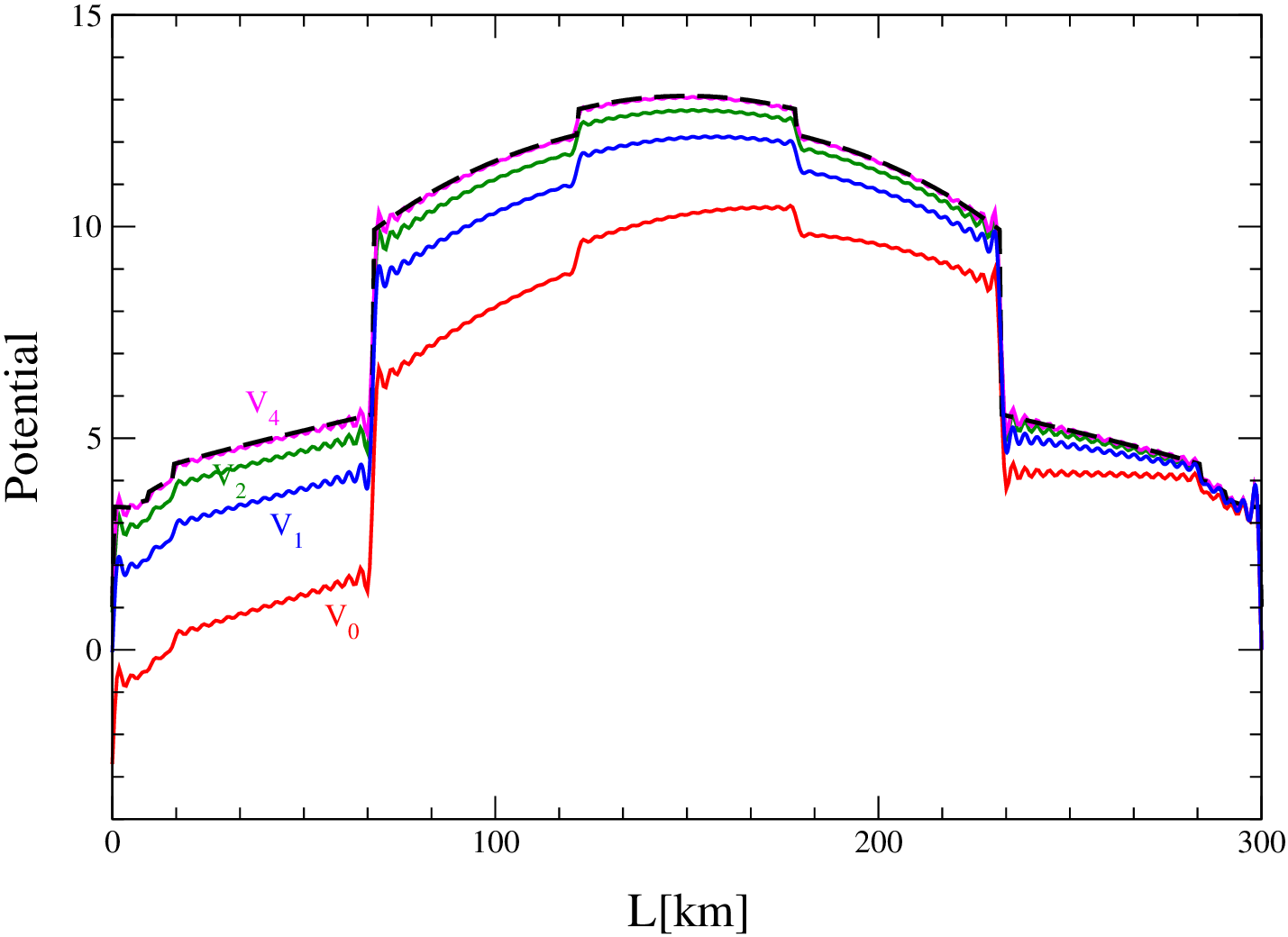}
  \caption{Same as in fig.\ref{fig:step-dmin1}, but for PREM-like profile 
(shown by dashed line), 
}  
 \label{fig:step-prem}
\end{figure}

\begin{figure}[htbp]
    \centering
    \includegraphics[width=.9\linewidth,clip]{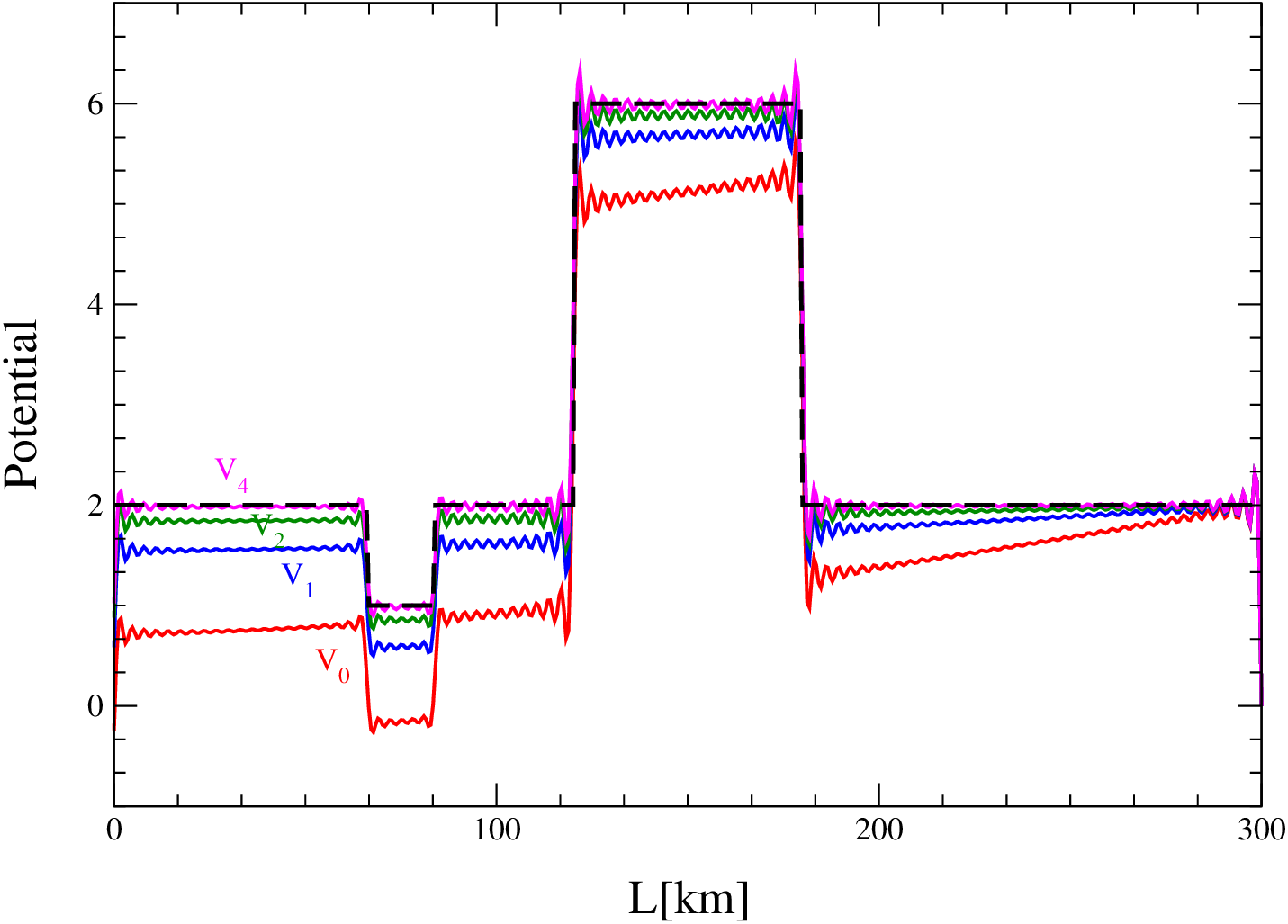}
    \caption{Same as in fig.~\ref{fig:step-dmin1}, but for asymmetric
    step-function potential (dashed line). } \label{fig:step-asym}
\end{figure}

This is illustrated in figs.~\ref{fig:step-dmin1} - \ref{fig:step-asym}. 
In fig.~\ref{fig:step-dmin1} we show the potential $V(x)$ 
for the step-function model of the Earth density profile, along with the 
zeroth-order reconstructed potential $V_0(x)$ and the results of the first, 
second and fourth iterations $V_1(x)$, $V_2(x)$ and $V_4(x)$ ($V_3(x)$ is 
not shown in order to avoid crowding the figure). The calculations were 
performed for $\delta_{min}=L^{-1}$, $\delta_{max}=300 L^{-1}$.  One can see 
that already the fourth iteration gives an excellent agreement with the exact 
potential. We have checked that for $\delta_{min}\ll L^{-1}$ already the 
zeroth-approximation potential $V_0(x)$ gives a very good accuracy (see 
also section~\ref{sec:small}). The wiggliness of the reconstructed potentials in 
fig.~\ref{fig:step-dmin1} is due to the finiteness of $\delta_{max}$; with 
increasing $\delta_{max}$ it decreases.

If $\delta_{min}$ exceeds the critical value (which for the chosen profile is 
approximately equal to $2.4 L^{-1}$ ), the successive iterations, instead of 
approaching the true potential, yield the potentials which more and more deviate 
from it. Thus, in this case the iteration procedure fails.  

The iteration procedure works very well not only for simple profiles like the 
step-function profile in fig.~\ref{fig:step-dmin1}, but also for more 
complicated ones. This is demonstrated in fig.~\ref{fig:step-prem}, which is 
similar to fig.~\ref{fig:step-dmin1}, but was plotted for the PREM-like density 
profile of the Earth \cite{Dziewonski:xy}. It should be stressed that this 
profile was used in fig.~\ref{fig:step-prem} for illustrative purposes only, 
since, in the form we used it, it is a realistic Earth density profile for 
$L=2R_\oplus \simeq 12742$ km and not for $L=300$ km (for which the Earth 
density is actually better approximated by the step-function profile of 
fig.~\ref{fig:step-dmin1}). 
Fig.~\ref{fig:step-asym} is similar to figs.~\ref{fig:step-dmin1} and 
\ref{fig:step-prem}, but was produced for an asymmetric Earth 
density profile (shown by the dashed line). It clearly demonstrates that 
asymmetric profiles can also be reconstructed very well, and so the 
inhomogeneities of matter distribution in the Earth can be studied by 
the method under consideration.   

Finally, we note that if $\delta_{max} L\gg 1$ but still not very large, one 
can devise an iterative procedure correcting for the corresponding error in the 
reconstructed potential (see Appendix \ref{app:Improved}). 

\subsubsection{Convergence properties of iterations}
\label{sec:conv}

We shall now discuss some properties of the iteration procedure
(\ref{V0m}) -- (\ref{iter1}). First, we note that the function $F(x,y;
2\delta_{min})$ is positive definite for all $0<x,y<L$ provided that
$\delta_{min}<2.246 L^{-1}$ (see Appendix B for the proof). One can
then show that for \begin{equation} \delta_{max} \to \infty~~~\mbox{and}~~~
\delta_{min}<{\rm min}\{2.246 L^{-1},\,\delta_1\}\,,
\label{cond1} 
\end{equation}
where $\delta_1$ will be defined shortly, the following string of
inequalities is satisfied:
\begin{equation}
V_0(x)<V_1(x)<V_2(x)<... <V(x)\,.
\label{ineq1}
\end{equation} 
This can be proven by induction. Consider first eq.~(\ref{inv2}).
{}From the positivity of $V(x)$ and $F(x,y; 2\delta_{min})$ it follows
that the second integral in (\ref{inv2}) (which we denote $I(x))$ is
positive, and so the first integral, which is the zeroth-approximation
potential $V_0(x)$, satisfies $V_0(x)<V(x)$. Generally, the larger
$\delta_{min}$, the smaller $V_0(x)$; for large enough $\delta_{min}$
the potential $V_0(x)$ may even become negative for some values of $x$
in the interval $[0,\,L]$. However, if $\delta_{min}$ does not exceed
certain limiting value $\delta_1$ (which we assume), the integral
$I_0(x)$ defined in eq.~(\ref{I0}) will still be positive. It is this
limiting value $\delta_1$ that appears in eq.~(\ref{cond1}). From the
definition of the integral $I_0(x)$ and the obtained condition
$V_0(x)<V(x)$ we then find $0<I_0(x)<I(x)$.  Therefore
$V_1(x)=V_0(x)+I_0(x)$ satisfies $V_0(x)<V_1(x)<V(x)$. Then, from the
definition of $I_1(x)$ we find $I_0(x)<I_1(x)<I(x)$, so that
$V_2(x)=V_0(x)+I_1(x)$ satisfies $V_1(x) <V_2(x)<V(x)$. Continuing
this procedure, we arrive at eq.~(\ref{ineq1}).

Thus, under the conditions of eq. (\ref{cond1}), each iteration produces the 
potential which is larger than that of the previous iteration. The potentials 
$V_n(x)$ approach the exact potential $V(x)$ from below and never exceed it. 
This is well illustrated by figs.~\ref{fig:step-dmin1} - \ref{fig:step-asym} 
(we recall that the wiggliness of the curves disappears in the limit 
$\delta_{max} \to\infty$). Conditions (\ref{cond1}) are thus sufficient for 
the convergence of the iteration procedure 
\footnote{It can be shown that the iteration potentials converge uniformly 
to $V(x)$ even for larger values of $\delta_{min}L$. Indeed, a sufficient 
condition for the uniform convergence is $\int_0^L F^2(x,y; 2\delta_{min})\,dx
\,dy < \pi^2$ (see, e.g., \cite{Goursat}), which is satisfied for 
$\delta_{min}L < 2.34$.}.

How can one estimate the critical value of $\delta_{min}$ 
above which the iteration procedure would diverge? From the preceding 
discussion it follows that the convergence of the iterations to the exact 
potential relies on the positivity of the ``correction terms'' $I_n(x)$. For 
values $\delta_{min}> \delta_1$ the integral $I_0(x)$ will become negative for 
some values of $x$ in the interval $[0,\,L]$, and so for those values of $x$ 
the potential $V_1(x)$ will deviate from $V(x)$ more than $V_0(x)$ does. 
If this propagates  into the further iterations, the iteration procedure 
would fail. However, it is possible that even if some $I_k(x)$ are not 
positive definite, higher-order correction integrals $I_n(x)$ with $n>k$ 
are. In this case the iteration procedure would still be convergent. 

In practice, it is difficult to determine the critical value of $\delta_{min}$ 
precisely. The closer (from below) $\delta_{min}$ to its critical value 
$\delta_{crit}$, the larger the number of iterations which is necessary to 
achieve a given accuracy of the reconstructed potential. Therefore, with 
$\delta_{min}$ approaching $\delta_{crit}$ it becomes more and more 
difficult to check if the procedure would converge. In addition, for 
$\delta_{min}$ close to $\delta_{crit}$ the behaviour of the iteration 
potentials $V_n(x)$ with increasing $n$ does not depend much on whether 
$\delta_{min}<\delta_{crit}$ or $\delta_{min}>\delta_{crit}$, so that it is 
difficult to decide if the critical value has already been exceeded. 
It is therefore reasonable to adopt some ``practical'' definition of 
$\delta_{crit}$, for example, as a value of $\delta_{min}$ for which the 
number of iterations necessary to reach a 10\% accuracy of the reconstructed 
potential reaches 100. For all the density profiles that we studied this 
value turned out to be $\delta_{crit}\simeq 2.4 L^{-1}$. We elaborate further 
on the convergence of the iterations in the next subsection.

Let us now return to the question of why the zeroth order potential
$V_0(x)$ correctly reproduces the positions and magnitudes of the
jumps in the exact potential $V(x)$. In eq.~(\ref{inv2}) (which is
exact in the linear regime), the second term on the right-hand side is
the integral over $y$ of the product of the continuous function
$F(x,y; 2\delta_{min})$ and the exact potential $V(y)$, which may have
discontinuities, but no $\delta$-function type singularities. Hence,
this integral is a continuous function of $x$. This immediately means
that all possible jumps in $V(x)$ are contained in the first integral
in eq.~(\ref{inv2}), i.e. $V_0(x)$.

\subsubsection{The limit of small $\delta_{min} L$}
\label{sec:small}

It is very instructive to consider the iteration procedure in the limit 
$\delta_{min} L\ll 1$. Since $x,y\le L$, the expression for $F(x,y; 
2\delta_{min})$ in this case simplifies to 
\begin{equation}
F(x,y; 2\delta_{min})\simeq 
\frac{16}{3}\,\delta_{min}^3 \,(L-x)(L-y)\,.
\label{F2}
\end{equation}
Then from eq.~(\ref{iter1}) we find a very simple result:
\begin{equation}
I_n(x)\simeq \frac{16}{3\pi}\,\delta_{min}^3 \,(L-x)\int_0^L V_n(y) (L-y) dy\,.
\label{In1}
\end{equation}
It means that the ``correction terms'' $I_n(x)$, which compensate for 
$\delta_{min} \ne 0$ in the inverse Fourier transformation, have a very 
simple coordinate dependence $\propto (L-x)$ for all $n$ and scale with 
$\delta_{min} L$ as $(\delta_{min} L)^3$. The same is true for the ``exact''
correction term $I(x)$ (the second integral in eq.~(\ref{inv2})), which is 
obtained from eq.~(\ref{In1}) by replacing in the integrand $V_n(y)$ by the 
exact potential $V(y)$. The approximately linear coordinate dependence of the 
deviation of the iteration potentials from the exact one can be seen even for 
$\delta_{min} L\sim 1$ (see figs.~\ref{fig:step-dmin1} - \ref{fig:step-asym}). 
It explains, in particular, why the deviations of $V_n(x)$ from the exact 
potential are small at $x\simeq L$ and largest at $x=0$.

Using eqs.~(\ref{iter1}) and (\ref{F2}), it is easy to show by 
induction that in the limit $\delta_{min} L\ll 1$ 
\begin{equation}
I_n(x)\simeq \frac{3}{2}\,v_0 \left(\frac{16}{9\pi}\,\delta_{min}^3 
L^3\right) 
\left[1-\left(\frac{16}{9\pi}\,\delta_{min}^3 L^3\right)^{n+1}\right] 
\,\frac{L-x}{L}\,,
\label{In2}
\end{equation}
\begin{equation}
V_n(x)\simeq V(x)-\frac{3}{2} v_0 \left(\frac{16}{9\pi}\,\delta_{min}^3 
L^3\right)^{n+1}\frac{L-x}{L}\,,~ ~~~~~~~~~~~~~~~~~
\label{Vn}
\end{equation}
where the constant $v_0$ is defined as
\begin{equation}
v_0=\frac{2}{L^2}\int_0^L V(y) (L-y) dy\,.
\label{V0}
\end{equation}
Note that these equations are in accord with eq.~(\ref{In1}). In the case 
of matter of constant density $V(x)=C_0=const$, one has $v_0=C_0$, and 
eq.~(\ref{Vn}) takes an especially simple form.   

{}From eq.~(\ref{Vn}) it follows that for $\delta_{min} L\ll 1$ the 
deviation of the $n$th-iteration potential $V_n(x)$ from the exact one 
scales as $[(16/9\pi)\,\delta_{min}^3 L^3]^{n+1}$, i.e. the convergence of 
$V_n(x)$ to the exact potential is very fast. 

Although eqs.~(\ref{In2}) and (\ref{Vn}) were obtained for $\delta_{min} L\ll 
1$, one can expect that they give correct order of magnitude estimates even 
for $\delta_{min} L\sim 1$. This allows one to estimate the number 
of iterations which is necessary to achieve a given accuracy of the 
reconstructed potential in the whole range $\delta_{min}\le \delta_{crit}$. 
First, from eq.~(\ref{Vn}) we find that in the 
case under consideration the critical value of $\delta_{min}$, above which 
the iteration procedure diverges, is $\delta_{crit}=(9\pi/16)^{1/3} L^{-1}$, 
so that (\ref{Vn}) can be rewritten as 
\begin{equation}
V_n(x)=V(x)-(3v_0/2)(\delta_{min}/\delta_{crit})^{3(n+1)} (L-x)/L\,. 
\label{Vn2}
\end{equation}
Assume now that we want the relative error in the reconstructed potential 
(which we define here as $|V(x)-V_n(x)|/v_0$)  
to be below $\varepsilon$. Then from (\ref{Vn2}) we find that the 
necessary number of iterations is 
\begin{equation}
n\simeq \frac{\ln(2\varepsilon/3)}{3\ln(\delta_{min}/\delta_{crit})}-1\,.
\label{n}
\end{equation}
As an example, take $\varepsilon=0.01$. Then for 
$\delta_{min}/\delta_{crit}=0.3$ already the zeroth-approximation potential 
$V_0(x)$ has the desired accuracy; for $\delta_{min}/\delta_{crit}=0.9$ 
eq.~(\ref{n}) gives $n\simeq 15$, for $\delta_{min}/\delta_{crit}=0.99$ it 
gives $n\simeq 165$, and in the limit $\delta_{min}\to\delta_{crit}$ one 
finds $n\to \infty$, as expected. It should be remembered, however, that for 
$\delta_{min}/\delta_{crit} \sim 1$ eq.~(\ref{n}) gives only a rough 
estimate of the necessary number of iterations because eq.~(\ref{Vn2}) was 
obtained in the limit $\delta_{min} L \ll 1$, which essentially coincides 
with $\delta_{min}/\delta_{crit} \ll 1$.

We have considered in this subsection the reconstruction of the potential 
$V(x)$ in the limit of small $\delta_{min} L$ iteratively only in order to 
illustrate some general features of the iteration procedure. In fact, 
for $\delta_{min} L\ll 1$ it is easy to find the closed-form solution of 
eq.~(\ref{inv2}) \footnote{Eq.~(\ref{inv2}) is a linear Fredholm integral 
equation of the second kind with the symmetric kernel $F(x,y;2\delta_{min})$. 
In the limit $\delta_{min} L\ll 1$ the kernel becomes separable (see 
eq.~(\ref{F2})), and the equation is trivially solved.}. Indeed, since in 
this case $I(x)\propto (L-x)$, the potential $V(x)$ can be written as 
$V(x)=V_0(x)+C_1 (L-x)$ with $C_1$ a constant. Substituting this into 
eq.~(\ref{inv2}), one readily determines $C_1$, which gives 
\begin{equation}
V(x)\simeq V_0(x)+C_2 \frac{\frac{16}{3\pi}\delta_{min}^3 L^3}
{1-\frac{16}{9\pi}\delta_{min}^3 L^3}\,\frac{L-x}{L}\, 
\label{V}
\end{equation}
with  
\begin{equation}
C_2=\frac{1}{L^2}\int_0^L V_0(y)(L-y) dy \,.
\label{C2a}
\end{equation}
Since the function $V_0(x)$ is known, the problem is solved.

\subsection{Iteration procedure II}
\label{sec:iterII}

\begin{figure}[htbp]
  \centering
  \includegraphics[width=.9\linewidth,clip]{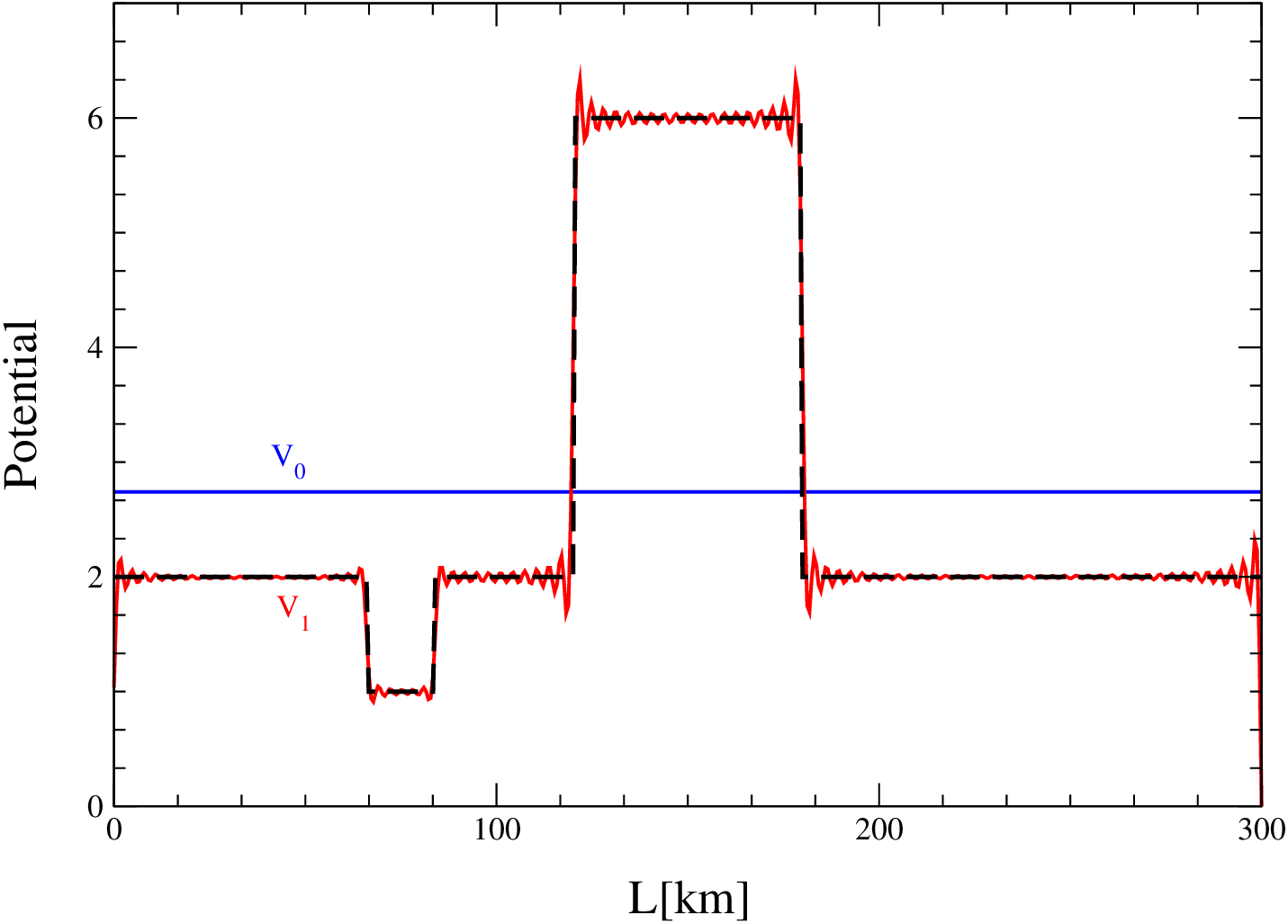}
  \caption{Asymmetric step-function profile (dashed line), zeroth 
  approximation potential $V_0(x)=\overline{V}$ and first order
  iteration potential $V_1(x)$. Vertical scale and values of 
  parameters are the same as in fig.~\ref{fig:step-dmin1}.} 
  \label{fig:iterII}
\end{figure}

In the iteration procedure considered in section \ref{sec:iterI} we
assumed that nothing is known \mbox{\em a priori} about the Earth
density profile, and the only experimental data available were the
neutrino data.  If some (even very rough) prior knowledge of the
matter density distribution inside the Earth exists, one can employ a
much better and faster iteration procedure to reconstruct the Earth
density profile. As an example, we consider here the case when the
average matter density along the neutrino path is known. The
corresponding average potential 
\begin{equation} \overline{V} = \frac{1}{L}
\int_0^L V(x)\,dx\,.
\label{Vbar}
\end{equation} 
is therefore known as well. 
The new iteration procedure is again described by eqs.~(\ref{I0}) 
- (\ref{iter1}), but the zeroth order approximation potential is now 
$V_0(x)=\overline{V}=const$. In fig.~\ref{fig:iterII} we plot this potential 
(horizontal line) and the first iteration potential $V_1(x)$ for the 
asymmetric step-function profile of fig.~\ref{fig:step-asym} (shown by the 
dashed line). One can see that, although $V_0$ is completely structureless, 
already the first iteration reproduces the exact profile extremely well. This 
should be compared with the results of iteration scheme I, for which only 
the fourth iteration gives similar accuracy (see fig.~\ref{fig:step-asym}).

To understand why this iteration scheme works so well, it is instructive to 
consider it in the limit $\delta_{min} L\ll 1$. First, we note that 
for an arbitrary $\delta_{min}$ and $n\ge 1$ eqs.~(\ref{inv2}) and 
(\ref{iter1}) yield 
\begin{equation}
V(x)-V_n(x) = \frac{1}{\pi}\int_0^L [V(y)-V_{n-1}(y)] 
F(x,y; 2\delta_{min}) 
dy\,.
\label{iter1a}
\end{equation}
In the limit $\delta_{min} L\ll 1$, the function $F(x,y; 2\delta_{min})$ is 
given by (\ref{F2}); substituting this into (\ref{iter1a}), for $n=1$ 
we find 
\begin{equation}
V(x)-V_1(x)\simeq \left(\frac{8}{3\pi}\,\delta_{min}^3 
L^3\right)\frac{L-x}{L} (v_0-\overline{V})\,,~ ~~~~~~~~~~~~~~~~~
\label{V1}
\end{equation}
where $v_0$ was defined in (\ref{V0}). For $n$th iteration ($n\ge 1$)
we find by induction
\begin{equation}
V(x)-V_n(x)\simeq \frac{3}{2} \left(\frac{16}{9\pi}\,\delta_{min}^3 
~L^3\right)^n\frac{L-x}{L} (v_0-\overline{V})\,.~ ~~~~~~~~~~~~~~~~~
\label{VnII}
\end{equation}
Comparing this with (\ref{Vn}), we find that in iteration scheme II the 
difference $V(x)-V_n(x)$ contains one less power of $[(16/9\pi)\,
\delta_{min}^3 L^3$], but is proportional to $v_0-\overline{V}$ rather 
than to $v_0$. We shall show now that for symmetric density profiles
($V(x)=V(L-x)$) the difference $v_0-\overline{V}$ vanishes, and therefore 
already the first iteration reproduces the 
exact profile. Indeed, from eqs.~(\ref{V0}) and (\ref{Vbar}) we have 
\begin{equation}
v_0-\overline{V}=\frac{1}{L^2}\int_0^L V(y)[2 (L-y)-L]\,dy=
\frac{2}{L^2}\int_{-L/2}^{L/2} V(z) z \,dy\,,
\label{Vz}
\end{equation}
where in the last integral we changed the integration variable to $z=L/2-y$.
For symmetric density profiles, $V(z)$ is an even function of $z$, and the 
integral in (\ref{Vz}) vanishes. 

What if the density profile is not symmetric? In that case one 
can write $V(z)=V_s(z)+V_a(z)$, where $V_s(z)$ and $V_a(z)$ are the symmetric 
and antisymmetric parts of $V(z)$. The symmetric part $V_s(z)$ does not 
contribute to the integral in (\ref{Vz}), whereas the contribution of $V_a(z)$ 
is small because the inhomogeneities of the Earth density distribution, 
responsible for $V_a$, occur on short coordinate scales \footnote{Generally, 
the contribution of $V_a$ to $v_0-\overline{V}$ is of order $\Delta V x_1/L$, 
where $\Delta V$ is the magnitude of the asymmetric inhomogeneity of the 
potential, and $x_1$ is the coordinate scale of the inhomogeneity. While 
$\Delta V$ may be of order of $V$, $x_1/L$ is always $\ll 1$.}. 
Thus, in scheme II, even for asymmetric density profiles the first iteration 
gives a very good approximation to the exact profile (see 
fig.~\ref{fig:iterII}).

It should be noted that, while the above discussion referred to the case 
$\delta_{min}L \ll 1$, fig.~\ref{fig:iterII} actually corresponds to 
$\delta_{min} L=1$. However, our consideration approximately applies to that 
case as well. For $\delta_{min} L\sim 1$, the integrand of the driving term for 
the second iteration in scheme II will be proportional, instead of $V(y)
(L/2-y)$, to the product of $V(y)$ and the function $[F(x,y; 2\delta_{min})-
(1/L)\int_0^L F(x,x'; 2\delta_{min})dx']$. This function, though not exactly 
antisymmetric with respect to $y\to L-y$, is almost antisymmetric (for all 
$x\in [0, L]$). This explains why iteration scheme II works so well even in 
the case $\delta_{min} L \sim 1$.

\section{Non-linear regime} 
\label{sec:nonlin}

Our previous discussion of the reconstruction of the Earth density
profile was based on the simple formula (\ref{f2}) for the function
$f(\delta)$, which was derived under the assumptions (\ref{cond2}).
The second of these conditions led to a rather stringent upper limit
on the neutrino path lengths in the Earth (\ref{upper}), or $L\lesssim
300$ km.  In order to be able to reconstruct the potential $V(x)$ over
larger distances, one has to employ an inversion procedure based on
the more accurate expression (\ref{f1}), which only requires the first
condition in (\ref{cond2}) for its validity. Let us now discuss this
improved expression.

Since the function $\omega(x)$ in eq.~(\ref{f1}) depends on $V(x)$, we face a 
non-linear problem now. The first condition in (\ref{cond2}) and the 
non-linearity condition $V L\gtrsim 1$ imply $\delta_{min} L\gg 1$. This, in 
turn, means that the matter density profile cannot be found by invoking an 
iteration procedure
\footnote{In sec.~\ref{sec:lin} we showed that in the linear regime
  the critical values of $\delta_{min}$, above which the iteration
  approach fails, correspond to $\delta_{min} L={\cal O}(1)$. The
  situation does not improve in the non-linear regime.}, and one
should resort to different methods of solving eq.~(\ref{f1}).  The
simplest possibility would be to discretize the problem and reduce the
non-linear integral equation (\ref{f1}) to a set of non-linear
algebraic equations, which can then be solved numerically.  However,
eq.~(\ref{f1}) is a non-linear Fredholm integral equation of the first
kind \footnote{We recall that integral equations of the first kind
  involve an unknown function only under the integration sign, whereas
  in equations of the second kind it is also present outside the
  integral.}, and equations of this type are notoriously difficult to
solve. Integral equations of the first kind belong to the so-called
ill-posed problems: their solutions are very unstable, and to arrive
at a reliable result one has to invoke special regularization
procedures. For linear integral equations, such procedures are well
developed (see section \ref{sec:finiteE} and Appendices C and D for
more details); however, no universal regularization techniques exist
for non-linear integral equations of the first kind. The situation is
further complicated by the fact that for $\delta_{min} L\gg 1$ the
potential $V(x)$ enters into the integrand of eq.~(2.2) being
multiplied by a fast oscillating function of the coordinate.  All this
makes the non-linear regime of the inverse problem of neutrino
oscillations in matter very difficult to explore. This regime
therefore requires a dedicated study, which goes beyond the scope of
the present paper.

\section{Experimental considerations}
\label{sec:exptl}

Up to now we ignored completely the experimental questions, such as the 
effects of the errors in experimental data and of finite energy resolution of 
neutrino detectors on the accuracy of the reconstructed matter density 
distributions. We now turn to these issues. We shall discuss here only the 
linear regime of the density profile reconstruction; the experimental 
questions pertaining to the non-linear regime will be considered 
elsewhere.

\subsection{Effects of experimental errors}
\label{sec:errors}

For solar neutrinos, the night-day asymmetry of the signal can 
be written as \cite{Akhmedov:2004rq}
\begin{equation}
A_{ND}=2\frac{N-D}{N+D}\simeq -c_{13}^2 \left[\frac{\sin^2 2\theta_{12}\,
\overline{\cos 2\hat{\theta}_{12}}}{1+\cos 2\theta_{12}\,
\overline{\cos 2\hat{\theta}_{12}}}\right] (f(\delta)/c_{13}^2)\,,
\label{And}
\end{equation}
where $\overline{\cos 2\hat{\theta}_{12}}$ is cosine of twice the effective 
1-2 mixing angle in matter, averaged over the neutrino production coordinate 
inside the Sun \cite{Blennow:2003xw}. The quantity $f(\delta)/c_{13}^2$ is 
independent of $c_{13}^2$ in the linear regime (see eqs.~(\ref{f2}) and 
(\ref{Vcc})). Thus, the relative error of the experimentally determined value 
of $f(\delta)/c_{13}^2$ is the sum of the relative errors of $A_{ND}$, 
$c_{13}^2$ and of the $\theta_{12}$ - dependent expression in the square 
brackets in eq.~(\ref{And}). The dependence of the latter on the error of 
$\theta_{12}$ is somewhat involved (mainly, because of the $\theta_{12}$ 
dependence of the effective mixing angle in matter $\hat{\theta}_{12}$). 
It can be approximated by 
\begin{equation}
\epsilon_{\theta_{12}}\simeq \left[1.74-\frac{1.83\,[1-(0.4-V_{av}/2\delta)^2]}
{(0.4-V_{av}/2\delta)[1+0.4(0.4-V_{av}/2\delta)]}\right] \Delta \theta_{12}\,,
\label{errtheta}
\end{equation}
where $V_{av}$ is the averaged over the neutrino production region value of 
the matter-induced neutrino potential in the Sun. The values of $V_{av}$ for 
various components of the solar neutrino spectrum can be found in table 1 
of ref.~\cite{deHolanda:2004fd}. 

The next question is how the error in the experimentally determined quantity 
$f(\delta)/c_{13}^2$ affects the accuracy of the reconstructed electron 
number density profile $N_e(x)$. Since, in the linear regime, to obtain a 
solution we invoke an iteration procedure, one might expect that the 
corresponding errors in the reconstructed density profile would accumulate 
with increasing iteration order. This is, however, not the case: at each 
iteration,  the relative error of the solution is the same as the relative error 
of $f(\delta)/c_{13}^2$, and the same applies to the exact profile $N_e(x)$. 
This follows from the fact that each iteration profile and the exact solution 
depend linearly on $f(\delta)/c_{13}^2$, which is a consequence of the linearity 
of eq.~(\ref{inv2}).

The main contribution to the error of $f(\delta)/c_{13}^2$ (and thus, of the 
reconstructed electron number density profile) is expected to come from the 
error in $A_{ND}$, which is by far the largest one (at the moment, more than 
100\%). Therefore, an accurate reconstruction of the matter density 
distribution inside the Earth would require very large detectors, capable of 
measuring the solar neutrino day-night effect with an accuracy commensurate 
with the desired accuracy of the matter density reconstruction. 

{}From eqs.~(\ref{f1}) and (\ref{omega}) it follows that the errors in the 
parameter $\Delta m_{21}^2$ and in the energy scale of neutrino detectors go 
linearly to the shifts in the reconstructed coordinate.

\subsection{Finite energy resolution of detectors}
\label{sec:finiteE}

Let us consider now the effects of finite energy resolution of neutrino 
detectors. We shall be assuming that neutrinos are detected through a 
charged-current capture reaction, so that the energy of the emitted electron 
in the final state directly gives the energy of the incoming neutrino. The 
electron energy, however, is not exactly measured because of the finite energy 
resolution of the detector, characterized by the resolution function $R(T_e, 
T_e')$. Here $T_e$ and $T_e'$ are the observed and true electron kinetic 
energies, respectively. Because of the one-to-one correspondence between the 
electron and neutrino energies, the resolution function can be written 
directly in terms of the ``observed'' and true neutrino energies.  In our 
discussion it proved more convenient to use $\delta=\Delta m_{21}^2/4E$ 
instead of the neutrino energy $E$, therefore we will be considering the 
detector resolution functions expressed in terms of $\delta$ and $\delta'$. 
In many cases the resolution function can be approximated by a Gaussian 
\begin{equation}
R(\delta, \delta')=\frac{1}{\sqrt{2\pi}\,\sigma(\delta)}\,
\exp\left\{-\frac{(\delta-\delta')^2}
{2\sigma^2(\delta)}\right\}
\label{Gauss1}
\end{equation}
with $\delta$-dependent width, e.g., $\sigma(\delta)=k_0\delta$ or
$\sigma(\delta)=k_0\sqrt{\delta_{max}\delta}$. 

When finite energy resolution of detectors is taken into account, the function 
$f(\delta)$ in the expression for the Earth regeneration factor in 
eq.~(\ref{reg}) has to be replaced by 
\begin{equation}
{\cal F}(\delta)=\int_0^\infty R(\delta, \delta') f(\delta')\,d\delta'\,,
\label{smooth}
\end{equation}
which describes the smoothing of $f(\delta)$ with the resolution function. In 
order to proceed with the density profile reconstruction, we have first to extract 
$f(\delta')$ from the experimentally measured quantity ${\cal F}(\delta)$. 
Once this has been done, one can employ the procedures described in sections 
\ref{sec:lin} or \ref{sec:nonlin}. Thus, instead of one-step inverse problem, 
we now have a two-step one. 

\subsubsection{Finite energy resolution effects on the Earth regeneration 
factor}

Let us first 
discuss the physical effects of the finite energy resolution of detectors  
on the observed Earth matter effect. 
In ref. \cite{Ioannisian:2004jk} 
it was shown that in the simplified case of the box-shaped resolution function 
with the energy width $\Delta E$, eq.~(\ref{f1})  has to be replaced by
\begin{equation}
{\cal F}(\delta)\simeq \int_0^L V(x)\frac{\sin\Delta(\delta)(L-x)}{\Delta(\delta)
(L-x)}\sin\left[2\int\limits_x^{L}\!\omega(x')\,dx'\right]dx \,, 
\label{f1a}  
\end{equation}
where $\Delta(\delta)$ is the resolution width in terms of $\delta$: 
$\Delta(\delta)=\delta \cdot(\Delta E/E)$. Eq.~(\ref{f1a}) differs from 
(\ref{f1}) by the extra factor 
$\sin \Delta(\delta)(L-x)/[\Delta(\delta)(L-x)]$ in the integrand. This factor 
equals unity when $\Delta(\delta)(L-x)=0$ (which corresponds to perfect energy 
resolution or $x=L$), and quickly decreases with increasing $\Delta(\delta)
(L-x)$ (see fig.~\ref{fig:g}). Thus, finite energy resolution of detectors 
leads to an attenuation of the contributions to the Earth matter effect coming 
from the density structures which are far from the detector, the attenuation 
length (distance from the detector) being
\begin{equation}
l_{att}~\simeq~ \frac{1}{\delta}\, \frac{E}{\Delta E}~=~\frac{1}{\pi}\, 
l_{osc}(E)\, \frac{E}{\Delta E}\,.
\label{atten}
\end{equation}
{}From eq.~(\ref{atten}) it follows that for the attenuation to be negligible, 
i.e. for $l_{att}$ to exceed considerably $L$ for all energies of interest, 
one needs 
\begin{equation}
\delta_{max} L\,\ll \frac{E}{\Delta E}\,.
\label{cond3}
\end{equation}
This can be in conflict with the requirement of good coordinate resolution 
$\delta_{max} L\gg 1$, unless the energy resolution is extremely good. 
Thus, finite energy resolution worsens the coordinate resolution of the 
reconstruction procedure. 
If one completely ignores the fact that finite energy resolution of neutrino 
detectors modifies the observed matter effect and simply uses ${\cal F}(\delta)$ 
instead of $f(\delta)$ for the density profile reconstruction, then in the 
case when condition (\ref{cond3}) is not satisfied it is essentially the 
finite energy resolution and not the value of $\delta_{max}$ that determines 
the coordinate resolution of the inversion procedure. Increasing $\delta_{max}$ 
beyond the limit (\ref{cond3}) would not lead to any significant improvement 
of the coordinate resolution in that case. 

Eq.~(\ref{f1a}) exhibits a power-law attenuation of the matter density 
structures far from the detector in the case of box-shaped energy 
resolution function. However, in a more realistic case of the Gaussian  
energy resolution, an even stronger (exponential) attenuation results. 
To show that, we first recall that the function ${\cal F}(\delta)$ is measured 
in a finite interval $\delta\in[\delta_{min},\delta_{max}]$. Although the 
integral in eq.~(\ref{smooth}) extends over the whole interval $0\le \delta' 
<\infty$, in fact the contributions to this integral coming from the domains 
of $\delta'$ which are far outside the interval $[\delta_{min},\delta_{max}]$ 
are strongly suppressed because of the presence of the resolution function 
$R(\delta, \delta')$ in the integrand. In particular, if $\delta_{min}$ 
exceeds a few widths of the Gaussian resolution function (\ref{Gauss1}), one 
can formally extend the integration in eq.~(\ref{smooth}) to negative values 
of $\delta'$ without affecting noticeably the value of the integral. 
Extending the integration to the interval $-\infty < \delta' < \infty$, from 
eqs.~(\ref{smooth}), (\ref{Gauss1}) and (\ref{f1}) one readily finds
\begin{equation}
{\cal F}(\delta)\simeq \int_0^L V(x)\,e^{-2(L-x)^2\sigma^2(\delta)}\,\sin
\left[2\int\limits_x^{L}\!\omega(x')\,dx'\right] dx \,. 
\label{f1b}  
\end{equation}

\subsubsection{Undoing the smoothing}

We now turn to the question of how to reconstruct the electron number
density profile of the Earth from the Earth regeneration factor in the
presence of a well-known but finite energy resolution.  As we already
pointed out, if one ignores the fact the observed Earth regeneration
factor is modified by the finite energy resolution of the detector and
naively uses ${\cal F}(\delta)$ instead of $f(\delta)$ for the density
profile reconstruction, the coordinate resolution of the
reconstruction procedure will in general be very poor. We have checked
numerically that for $\delta_{max} L\gtrsim 100$ and the energy
resolution width $\Delta E/E=\Delta(\delta)/\delta$ exceeding 1\%, the
reconstructed profile differs sizeably from the true one. However, if
the resolution function is well known, one can do a much better job:
first, recover the function $f(\delta')$ from the experimentally
measured quantity ${\cal F}(\delta)$, and then reconstruct the Earth
density profile from $f(\delta')$.  In other words, one could to some
extent undo the smoothing of $f(\delta')$ caused by the finite energy
resolution of the neutrino detectors and described by
eq.~(\ref{smooth}). This is similar to the procedures one has to
invoke in various remote sensing problems (e.g., in medical
tomography).

In principle, if the resolution function $R(\delta, \delta')$  is exactly 
known and the function ${\cal F}(\delta)$ is precisely determined from the 
experiment, one could expect that the function $f(\delta')$ can be exactly 
found from (\ref{smooth}). This is, however, not the case. The point is 
that eq.~(\ref{smooth}) is a Fredholm integral equation of the first kind,
which belongs to the class of ill-posed problems: small variations in 
${\cal F}(\delta)$ can lead to very large changes in the reconstructed function 
$f(\delta')$. To illustrate this, let us note that for any integrable 
$R(\delta,\delta')$ and an arbitrary constant $A$
\[
\lim_{k\to\infty}\, \int_0^\infty R(\delta, \delta')\,A \sin(k
\delta')\,d\delta' ~=~0\qquad\quad
\]
by Riemann-Lebesgue lemma. Therefore, even for very large values of $A$,  
changing $f(\delta')\to f(\delta')+ A\sin(k\delta')$ practically does not 
modify the observable ${\cal F}(\delta)$ if $k$ is large enough. This means 
that small variations in ${\cal F}(\delta)$ correspond to very large 
high-frequency changes in the reconstructed function $f(\delta')$. The 
solution does not depend continuously on the data, i.e. is highly unstable. 

The function ${\cal F}(\delta)$ is always determined experimentally with some 
errors; the errors are drastically magnified in the process of solving   
eq.~(\ref{smooth}) for $f(\delta')$. Even if one assumes ${\cal F}(\delta)$ to 
be exactly known, the rounding errors, which are inherent to any numerical 
calculation, would play the same role as the experimental errors and 
destabilize the solution. Therefore, straightforward methods of solving 
Fredholm integral equations of the first kind (such as discretization and 
reduction to a system of linear algebraic equations) do not work. One has to 
employ some regularization in order to suppress high-frequency noise in the 
solution. 

There exist many regularization approaches for ill-posed problems and
a vast literature on the subject. In our calculations we use two
popular regularization schemes: the truncated singular value
decomposition (TSVD) \cite{TSVD} and the Backus-Gilbert (BG) method
\cite{BG,LE,num}, which are described in Appendices C and D,
respectively.  The results of calculations for the step-function
density profile and the Gaussian detector resolution function
(\ref{Gauss1}) with 10\% resolution ($k_0=0.1$) are presented in
figs.~\ref{fig:TSVD} and \ref{fig:BG}. In these figures shown are the
exact profile (dashed line) and the reconstructed profiles in the case
of Gaussian resolution: the naive calculation, which used ${\cal
  F}(\delta)$ instead of $f(\delta')$, is shown by the dash-dotted
curve, while the results of the corresponding regularization
approaches are shown by the solid curves. In the case of the TSVD
regularization, the contributions of the singular values $\lambda_i\le
10^{-10}$ were truncated (see Appendix C); in the BG approach,
1501-point calculation has been used. One can see from the figures
that, while the naive calculation gives very poor reconstruction of
the density profile far from the detector, undoing the smoothing
caused by the finite energy resolution of the detectors improves the
quality of the reconstruction drastically.

\begin{samepage}
\begin{figure}[tbp]
  \centering 
  \includegraphics[width=.9\linewidth,clip]{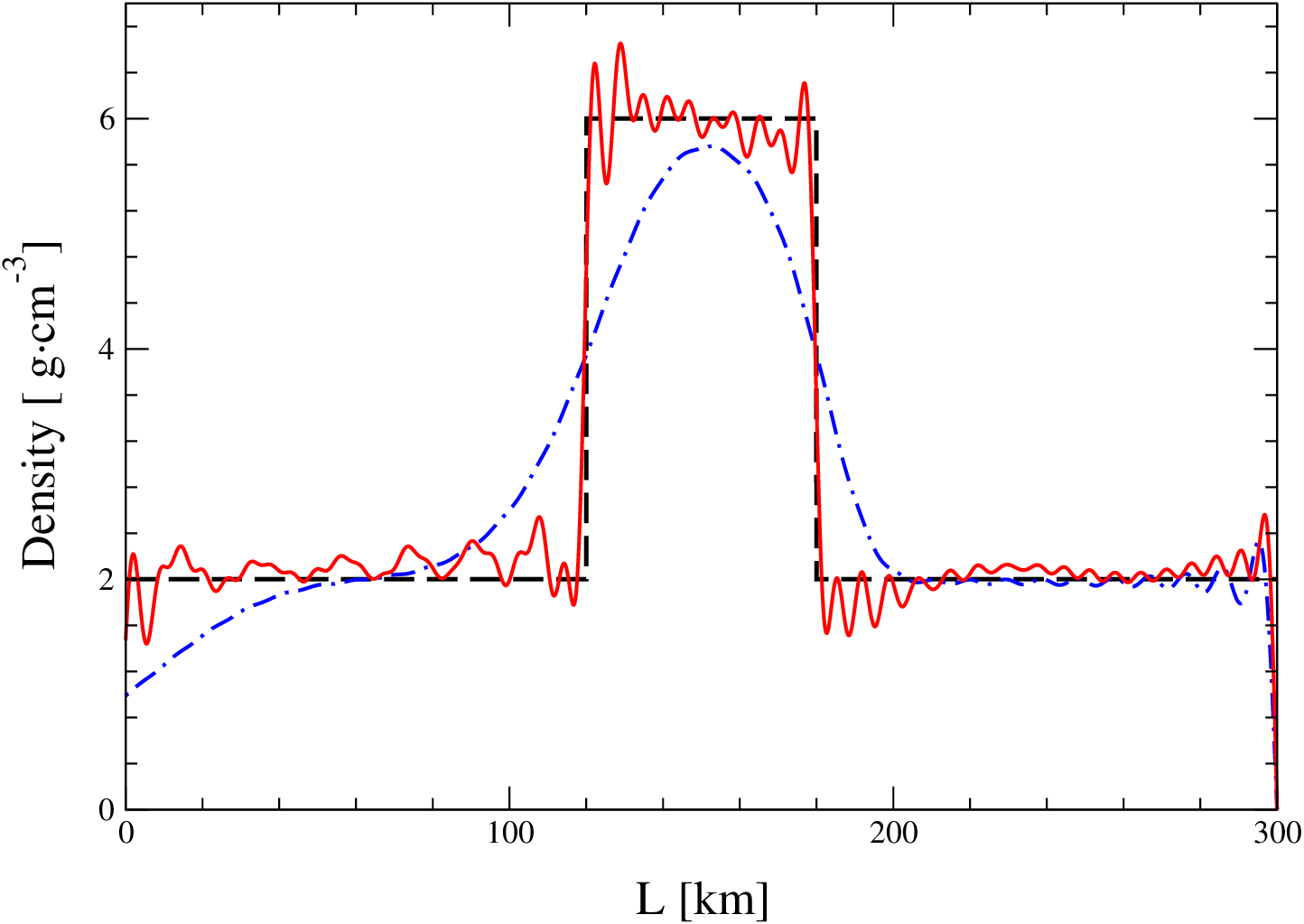}
  \caption{Density reconstruction for step-function density profile
  using TSVD method. The dashed line is the exact profile, the
  dash-dotted curve corresponds to a naive calculation using ${\cal
  F}(\delta)$ instead of $f(\delta')$ for 10\% Gaussian resolution,
  the solid curve is the result of TSVD regularization for Gaussian
  resolution, truncation at $\lambda_i\le 10^{-10}$.} \label{fig:TSVD}
\end{figure}
\begin{figure}[htbp]
  \centering 
  \includegraphics[width=.9\linewidth,clip]{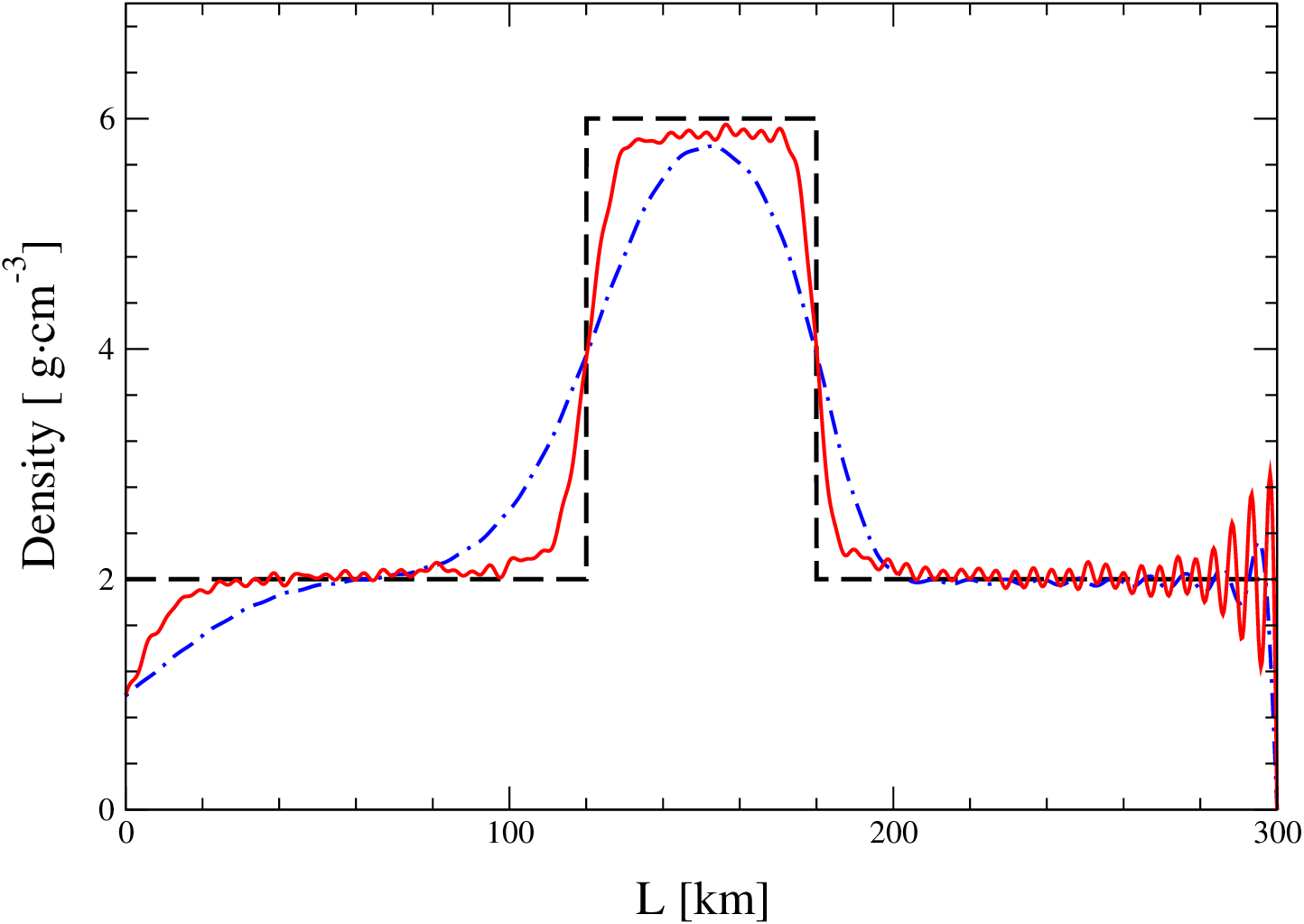}
  \caption{Same as in fig.~\ref{fig:TSVD}, but for Backus-Gilbert
  method with $N=1501$ points. } \label{fig:BG}
\end{figure}
\end{samepage}

Comparing figs.~\ref{fig:TSVD} and \ref{fig:BG} one can see that the BG 
method gives in general more accurate reconstruction of the profile in the 
regions that are far from the ends of the neutrino trajectory in the Earth, 
whereas the TSVD method better reproduces the 
exact profile near these endpoints, $x\simeq 0$ and $x\simeq L$, and the 
jumps of the density (at the jumps, the TSVD curve is steeper than that of 
the BG approach). One can employ both these methods of data treating for the 
same set of experimental data and thus combine the advantages of each  
approach.

\section{Discussion and outlook}
\label{sec:disc}

\vspace*{2mm} We have developed a novel, {\em direct} approach to
neutrino tomography of the Earth, based on a simple analytic formula
for the Earth matter effect on oscillations of solar and supernova
neutrinos inside the Earth.  These neutrinos have a number of
advantages over the accelerator neutrinos from the viewpoint of the
tomography of the Earth. Apart from coming from a free neutrino
source, they are sensitive to the Earth matter effects even if the
distances they travel in the Earth are as short as $\sim 50$ - 100 km.
In addition, they can be used for reconstructing asymmetric density
profiles and thus are sensitive to short scale inhomogeneities of
matter distribution in the Earth, as possibly relevant in prospection
applications.  Neutrino tomography allows the reconstruction of the
electron number density of the Earth, i.e. is sensitive to its
chemical composition. In this respect it can nicely complement
geoseismological studies, which are based on the measurements of the
spatial distributions of seismic wave velocities and can only probe
the total density of the Earth's matter, but not its chemical
composition.

In the linear regime, which is valid for relatively short distances
traveled by neutrinos in the Earth (up to several hundred km), the
problem amounts to finding an inverse Fourier transform of the Earth
regeneration factor.  However, to perform this transformation, one
needs to know the Earth regeneration factor in the whole infinite
interval of neutrino energies $0\le E < \infty$, whereas
experimentally it can only be measured in a finite range $E_{min}\le
E\le E_{max}$. Non-vanishing $E_{min}$ leads to a finite coordinate
resolution $\Delta x\sim l_{osc}(E_{min})$ of the reconstruction
procedure, while finite $E_{max}$ results in a systematic shift of the
reconstructed profile compared to the true one.

Future experiments should be able to detect solar neutrinos with
energies as small as $\sim 100$ keV, which means that a very good
coordinate resolution can in principle be achieved; at the same time,
finite $E_{max}$, i.e.  the lack of knowledge of the Earth
regeneration factor in the high energy domain, could potentially be a
serious drawback for the neutrino tomography of the Earth. We have
shown, however, that this ignorance of the Earth matter effect at high
neutrino energies can be compensated for by making use of simple
iterative procedures. We have developed two such iteration schemes,
one of them not using any prior information about the Earth density
profile, and the other assuming the average matter density along the
neutrino path to be known. Both schemes give very good results, the
second one leading to a faster convergence.

In order to reconstruct the Earth matter density profiles over the
distances of geophysical interest, i.e. exceeding a few hundred km,
one needs to solve a complex non-linear problem. It should be noted,
however, that even in the linear regime ($L\lesssim\,$ a few hundred
km) one can obtain very interesting information about the structure of
the Earth's crust and upper mantle. The maximal depth $d$ of the
neutrino trajectory inside the Earth is \begin{equation}
d=R_\oplus-\sqrt{R_\oplus^2-(L/2)^2}\,,
\label{depth} 
\end{equation}
where $R_\oplus=6371$ km is the average radius of the Earth. For $L=300$ km 
this gives $d\simeq 1.8$ km,  which could be quite sufficient, e.g., for 
minerals, oil and gas prospecting. Learning more about the large-scale 
structure of the Earth's interior would require calculations in the non-linear 
regime. 

There is a number of very interesting physics and mathematics issues
related to the problem of neutrino tomography of the Earth, which
still are to be explored. We have studied in detail the linear regime
of the Earth density profile reconstruction; attacking the non-linear
regime remains a challenge for future investigations.
Another issue is recovering the true Earth regeneration factor,
necessary for the inversion procedure, from the experimentally
measured one, which is smoothed due to the finite energy resolution of
the detector. Similar procedures are invoked in various remote sensing
problems (e.g., in medical tomography, early vision, inverse problems
of geo- and helioseismology, etc.).  We have used two relatively
simple methods of undoing the smoothing effects, the truncated
singular value decomposition and the Backus-Gilbert method, and only
in the linear regime. At the same time, there is a variety of other
methods developed for solving problems of this kind; it would be
interesting to study some of them in application to the neutrino
tomography of the Earth, and also to consider the de-smoothing
procedure in the non-linear regime.

Due to the rotation of the Earth, for a fixed position of the detector
solar neutrinos will probe the Earth density distribution along a
number of chords, ending at the detectors and spanning a segment
inside the Earth. The size of the segment will depend on the
geographic location of the detector. Because of the relatively low
statistics, only density distributions averaged over certain angular
bins of incoming neutrino directions will be probed. In order to
reduce these averaging effects, i.e. to have small angular bins, one
would need very large detectors and/or large overall detection time.
Supernova explosions being one-shot events, the Earth tomography with
supernova neutrinos will allow the density profile reconstruction only
along a single neutrino path for each detector and each supernova,
depending on the detector and supernova positions at the time the
supernova neutrinos arrive at the Earth.

The accuracy of the reconstruction of the Earth matter density
distribution will depend crucially on the accuracy of the experimental
data -- of the measured Earth regeneration factor and of the neutrino
oscillation parameters. It will also depend on the accuracy of the
reconstruction procedure itself. The largest error is expected to come
from the errors in the measured Earth matter effect, mainly because
this effect is rather small for solar and supernova neutrinos. To
achieve a given accuracy of the density profile reconstruction, one
should measure the Earth matter effect with similar or better
accuracy. The present value of the night-day asymmetry of the solar
neutrino signal, measured at Super-Kamiokande, is $2.1 \% \pm 2.0\%
(\rm{stat}) \pm 1.3 \% (\rm{syst})$~\cite{Smy:2003jf}, i.e. the error
constitutes more than 100\%. This is certainly a very important
limiting factor for the reconstruction method described here.  Future
megaton water Cherenkov detectors, such as UNO or Hyper-Kamiokande,
are expected to reduce this error by a factor of 4 to 10; however,
such detectors, unfortunately, are not very suitable for neutrino
tomography of the Earth. The reason is that they are based on the $\nu
e$ scattering process, in which the measured energy of the recoil
electron gives only loose information on the energy of the incoming
neutrino \footnote{In $\nu e$ scattering the neutrino
  energy could be determined if, in addition to the energy of the
  recoil electron, the direction of its momentum were measured.
  Precise determinations of these directions are not possible with
  water Cherenkov detectors, but can be realized in ICARUS-type liquid
  argon detectors \cite{Amerio:2004ze}. }.  Detectors based on
charged-current neutrino capture reactions, such as LENS or MOON
\cite{Raghavan,Ejiri:1999rk,LOWE}, are probably more promising. It is
not clear, however, whether sufficiently large (megaton or perhaps
even tens of megaton scale) detectors of this kind can be constructed,
and much work still has to be done to assess the feasibility of
neutrino tomography of the Earth.  In any case, studying the Earth's
interior with neutrinos is an exciting possibility, which can add yet
another motivation for very large detectors of solar and supernova
neutrinos.

\vspace*{4mm}
{\em Acknowledgements.} 
The authors are grateful to G. Fiorentini and A. Yu. Smirnov for useful 
discussions. This work was supported by Spanish grant BFM2002-00345 and by 
the European Commission Human Potential Program RTN network 
MRTN-CT-2004-503369. The work of E.A. was partially supported by the 
sabbatical grant No. SAB2002-0069 of the Spanish MECD.   M.A.T.\ is
supported by the M.E.C.D.\ fellowship AP2000-1953.

\begin{appendix}

\section{The case of not very large $\delta_{max} L$}
\label{app:Improved}

If $\delta_{max} L\gg 1$ but still not very large, one can develop an
iterative procedure correcting for the corresponding error in the
reconstructed potential.

Let us add to and subtract from eq.~(\ref{aux1}) the corresponding integral 
from $\delta_{min}$ to $\delta_0$, where $\delta_0\gg \delta_{max}$ is a 
``numerical emulation of infinity'', i.e. a large number for which one can 
neglect the difference between $\sin (\delta_0 x)/x$ and $\pi\delta(x)$ (e.g., 
$\delta_{0}=10^3 L^{-1}$). This gives 
\begin{eqnarray}
V(x)\simeq \frac{4}{\pi}\int_{\delta_{min}}^{\delta_{max}} 
f(\delta)\sin 2\delta\,(L-x)\,d\delta\, +\frac{1}{\pi}\int_0^L V(y)
\{F(x,y; 2\delta_{min})\qquad\quad \nonumber \\
+\,F(x,y; 2\delta_{0})-F(x,y; 2\delta_{max})\}\,dy\,.
\label{improved}
\end{eqnarray}
If eq.~(\ref{f2}) is exact, eq.~(\ref{improved}) becomes exact in the limit 
$\delta_0\to\infty$. Note that the first term on the right hand side of 
(\ref{improved}) contains integration over $\delta$ from $\delta_{min}$ to 
$\delta_{max}$, 
and not from $\delta_{min}$ to $\infty$. Thus, it properly takes into account 
that the function $f(\delta)$ is only measured in the energy interval 
$E_{min}\le E\le E_{max}$ with $E_{min}\ne 0$. 

Eq.~(\ref{improved}) can now be solved by iterations, the procedure 
being quite analogous to the one used for solving eq.~(\ref{inv2}) in sections 
\ref{sec:iterI} or \ref{sec:iterII}. If $\delta_{max} L$ is a very large 
number, the difference between the solutions of eqs. (\ref{inv2}) and 
(\ref{improved}) is numerically insignificant, but it can be noticeable when 
$\delta_{max} L$ is not too large. In our calculation in the present paper we 
always use $\delta_{max}L\ge 100$, so that there is no need to solve the 
improved equation (\ref{improved}).

\label{app:derivation}
\renewcommand{\theequation}{\thesection\arabic{equation}}
\setcounter{equation}{0}
\section{Properties of $F(x,y; 2\delta_{min})$}
\label{sec:properties}

Let us prove that 
\begin{equation}
F(x,y; 2\delta_{min})>0 \quad \mbox{for all}\quad 0 < x,y <L\quad 
\mbox{and}\quad \delta_{min}<2.246 L^{-1}\,. 
\label{Fprop1}
\end{equation}
Indeed, the function $F(x,y; k)$ defined in eq.~(\ref{F1}) can be written as 
$F(x,y; k)=g(x'-y')-g(x'+y')$, where $g(z)=\sin kz/z$, $x'=L-x$, $y'=L-y$. For 
positive $k$, the function $g(z)$  reaches its first minimum at $kz\simeq 
4.493$ and is a decreasing function of $z$ for $0< kz < 4.493$ (see 
fig.~\ref{fig:g}). Since $|x'-y'|\le x'+y'$, $F(x,y; k)$ is positive for all 
$0<x,y<4.493/k$. Substituting $k=2\delta_{min}$ and requiring 
$0<x,y<L<4.493/2\delta_{min}$, one arrives at (\ref{Fprop1}).

\section{Truncated singular value decomposition (TSVD) }
\label{sec:TSVD}

Any square-integrable function $R(\delta, \delta')$ can be represented as 
\cite{TSVD} 
\begin{equation}
R(\delta,\delta')~=~\sum_{i=1}^\infty \lambda_i\,u_i(\delta)\,v_i(\delta')\,.
\qquad\quad
\label{C1}
\end{equation}
where $\lambda_i$ are the so-called singular values, which converge to 0 
as $i\to\infty$, and $u_i(\delta)$ and $v_i(\delta')$ are the singular 
functions. They are orthogonal and can be normalized to satisfy 
\begin{equation}
(u_i\cdot u_j)~\equiv~\int u_i(\delta) u_j(\delta)\,d\delta~=~\delta_{ij}\,,
\label{C2}
\end{equation}
and similarly for $v_i$. There exist standard programs for finding singular 
values and singular functions of square-integrable kernels.

{}From eq.~(\ref{smooth}), which can be written in symbolic form as 
${\cal F}=R\cdot f$, and eqs.~(\ref{C1}) and (\ref{C2}) one finds 
\begin{equation}
f(\delta')~=~\sum_{i=1}^\infty \frac{(u_i\cdot{\cal F})}{\lambda_i}\,
v_i(\delta')\,.\qquad\qquad
\label{TSVDres}
\end{equation}
Since $\lambda_i\to 0$ as $i\to\infty$, the contributions of higher 
harmonics to $f(\delta')$ are strongly enhanced, which leads to instabilities 
due to small variations in ${\cal F}(\delta)$. To suppress these instabilities 
(regularize the solution), one can truncate the series at certain value of 
$i$, i.e. remove the contributions of very small singular values:
\begin{equation}
f(\delta')~\simeq~\sum_{i=1}^N \frac{(u_i\cdot{\cal F})}{\lambda_i}\,
v_i(\delta')\,.\qquad\qquad
\end{equation}
In our calculations we truncate the series when $\lambda_i$ become smaller 
than $10^{-10}$. 

A variant of the TSVD method employs a soft truncation, in which in the sum 
in eq.~(\ref{TSVDres}) the following substitution is made: 
\begin{equation}
\frac{1}{\lambda_i}\to\frac{\lambda_i}{\lambda_i^2+\alpha^2}\,,\qquad\qquad
\end{equation}
where $\alpha$ is the regularization parameter. This procedure is equivalent 
to using the Tikhonov regularization \cite{TSVD}. 

\section{Backus-Gilbert method }
\label{sec:BG}

This method \cite{BG,LE,num} is especially well suited for incorporating the 
errors of experimental data into the inversion procedure.  

We want to solve eq.~(\ref{smooth}) for the function $f(\delta')$.
First, we take into account that the actual experiments provide us with 
the binned data, i.e. we will have not a function ${\cal F}(\delta)$, but 
a finite set of discrete values ${\cal F}(\delta_i)\equiv{\cal F}_i$ 
\mbox{($i=1,\dots,N$)}. 
{}From eq.~(\ref{smooth}) we then have 
\begin{equation}
{\cal F}_i=\int_0^\infty R(\delta_i, \delta') f(\delta')\,d\delta'\,.
\label{BG1}
\end{equation}
Then, we seek the solution of this equation in the form
\begin{equation}
f(\delta')~\simeq~ \sum_{i=1}^N a_i(\delta'){\cal F}_i\,,
\label{BG2}
\end{equation}
where the coefficients $a_i(\delta')$ are to be determined. This is, actually, 
the most general form for linear inversion. Substituting here ${\cal F}_i$ 
from eq. (\ref{BG1}), we find
\begin{equation}
f(\delta')\simeq \int_0^\infty \hat{\delta}(\delta', \delta'') 
f(\delta'') 
\, d\delta''\,,
\label{BG3}
\end{equation}
where the function $\hat{\delta}(\delta', \delta'')$ is given by
\begin{equation}
\hat{\delta}(\delta', \delta'')~=~\sum_{i=1}^N a_i(\delta') 
R(\delta_i,\delta'')\,.
\label{BG4}
\end{equation}
{}From eq. (\ref{BG3}) it is seen that in the ideal case the function 
$\hat{\delta}(\delta', \delta'')$ should coincide with Dirac's 
$\delta$-function. In reality it does not, but we can try to make it as close 
to the $\delta$-function as possible. The simplest Backus-Gilbert prescription 
for that is the following. Define the spread function $r(\delta')$ as
\begin{equation}
r(\delta')~=~ \int_0^\infty (\delta'-\delta'')^2 \,[\hat{\delta}(\delta', 
\delta'')]^2 \, d\delta''\,. 
\label{BG5}
\end{equation}
It is non-negative, and would vanish if the function $\hat{\delta}(\delta', 
\delta'')$ were indeed Dirac's $\delta$-function. We now find the coefficients 
$a_i(\delta')$ by minimizing $r(\delta')$, subject to the normalization 
constraint 
\begin{equation}
\int_0^\infty \hat{\delta}(\delta', \delta'') \, d\delta'' ~=~ 1\,.
\label{BG6}
\end{equation}
This would lead to $\hat{\delta}(\delta', \delta'')$ being closest to the 
$\delta$-function for a given choice of the spread\footnote{Other choices of 
the spread function are also possible.}. 
The derivation is straightforward, and here we give the results.

Let us define the matrix $S(\delta')$ as 
\begin{equation}
S_{ik}(\delta') ~=~ \int_0^\infty (\delta'-\delta'')^2 \, R(\delta_i, 
\delta'')\,R(\delta_k, \delta'') \, d\delta''\,. 
\label{BG7}
\end{equation}
We also define $N$ numbers $u_i$:
\begin{equation}
u_i ~=~\int_0^\infty R(\delta_i, \delta_1)\,d\delta_1\,.
\label{BG8}
\end{equation}
\noindent
If the energy resolution function $R(\delta, \delta')$ is normalized 
to unit integral, then all $u_i$ in eq.~(\ref{BG8}) are equal to 1, but 
in general $u_i$ can be different from unity. Minimization of the function 
$r(\delta')$, subject to the normalization constraint (\ref{BG6}), yields
\begin{equation}
a_i(\delta')~=~
\frac{\sum\limits_{k=1}^N [S^{-1}(\delta')]_{ik}\,u_k}
{\sum\limits_{j,k=1}^N u_j\,[S^{-1}(\delta')]_{jk}\, u_k}\,,
\label{BG9}
\end{equation}
or, in matrix notation,
\begin{equation}
a(\delta')~=~\frac{S^{-1}(\delta')\, u}{u\, S^{-1}(\delta')\,u}\,.
\label{BG10}
\end{equation}
Note that the matrix $S$ is symmetric and positive-definite, so it has an 
inverse.

Using eq. (\ref{BG2}), one then finds the approximate solution of 
eq.~(\ref{BG1}):
\begin{equation}
f(\delta')~\simeq~
\frac{\sum\limits_{i,k=1}^N {\cal F}_i\,[S^{-1}(\delta')]_{ik}\,u_k}
{\sum\limits_{j,k=1}^N u_j\,[S^{-1}(\delta')]_{jk}\, u_k}\,,
\label{BG11}
\end{equation}
or, using the matrix notation,
\begin{equation}
f(\delta)'~\simeq~\frac{{\cal F}\, S^{-1}(\delta')\, 
u}{u\, S^{-1}(\delta')\, u}\,.
\label{BG12}
\end{equation}

Some comments are in order. To have a good accuracy, one would like the 
number of data bins $N$ to be large, but then $S(\delta')$ is a very large 
matrix, and inverting it may be a time and memory consuming operation. It 
is, however, not necessary to invert $S(\delta')$. It is enough to solve the 
system of $N$ linear equations
\begin{equation}
\sum_{k=1}^N S_{ik}(\delta')y_k(\delta')=u_i
\label{BG13}
\end{equation}
and find an $N$-component vector $y(\delta')$, which is a much simpler 
problem. Then eqs.~(\ref{BG11}) and (\ref{BG12}) can be rewritten as
\begin{equation}
f(\delta')~\simeq~ \frac{\sum\limits_{i=1}^N {\cal F}_i\,y_i(\delta')}
{\sum\limits_{k=1}^N u_k\, y_k(\delta')}~=~
\frac{{\cal F} y(\delta')}{u\, y(\delta')}\,.
\label{BG14}
\end{equation}

\end{appendix}


\end{document}